\documentclass{ws-rv9x6}
\usepackage{subfigure}   
\usepackage{ws-rv-thm}   
\usepackage{ws-rv-van}
\usepackage{ws-index}   
\makeindex
\newindex{aindx}{adx}{and}{Author Index}       
\renewindex{default}{idx}{ind}{Subject Index}  

\begin{document}

\chapter[Some Old And New Puzzles In The Dynamics Of Fluids]{Some Old And New Puzzles\\ In The Dynamics Of Fluids\label{ch1}}

\author[Ihor Mryglod and Vasyl' Ignatyuk]{Ihor Mryglod and Vasyl' Ignatyuk 
}

\address{Institute for Condensed Matter Physics, National Academy of Sciences of Ukraine,\\
Svientsitskii str.~1, Lviv, Ukraine 79011, \\
mryglod@icmp.lviv.ua 
}

\begin{abstract}
One of the basic concepts of modern physics with a long prehistory is a fluid, which means a substance that flows under an applied shear stress. In this sense fluids form a wide subset of the phases of matter that includes liquids, dense gases, plasmas, and to some extent even plastic solids. Fluidity is one of the main dynamical characteristics that depends strongly on the details of the local structure. And vice versa, dominant details of local structural ordering in the arrangement of particles on some time scales are important for understanding the dynamics of fluids. The orderliness over distances comparable to the inter-particle distances is usually treated as the short-range order, whereas the orderliness repeated over infinitely large distances is called the long-range order. Both are absent in the ideal gas, but liquids and amorphous solids exhibit the short-range order. The physics of crystalline solids with the long-range order is well understood. In fluids, however, the atomic structure is changing with time. How are specific features of structural ordering reflected in the fluid dynamics? In this chapter we try to find answers to some old and new questions that still make the fluid dynamics still very attractive for the theoretical studies.
\end{abstract}


\body

\tableofcontents

\section{Introduction}\label{intro}

This Chapter is mainly based on the talk, delivered by one of us (I.~M.) at the Ising Lectures 2017. We do not deal directly with the critical phenomena, but discuss some problems of the dynamics of fluids, where the terms \textit{order} and \textit{disorder} play a significant role and \textit{cooperative phenomena} are very  important. In this context our goal is rather close to the mainstream of Ising Lectures. As stated in the book published recently in Ref.~\refcite{Ising20} on the occasion of the 20th anniversary of the Ising Lectures: 
\begin{quote}
``Nowadays Ising-like models completed by ideas from complex system science come into play as simple models on the way to \textit{conceive the secrets of nature in all their complexity}''.
\end{quote}
In the same way we shall try to describe the fluid dynamics by relatively simple models, which: i) are based on the microscopic foundations; ii) whenever possible, yield rigorous analytical expressions for the quantities of interest; iii) can be verified in the computer simulations. Realizing that even the most refined model is nothing more but the ``working tool'' for the researcher to explain the Nature's mysteries, and that it could render an eventual success only in a conjunction with a suitable theory (``there is nothing as practical as a good theory'', as a well known aphorism states\footnote{It is interesting that the origin of this aphorism was a subject of special investigation\cite{JMH}.}), we pay great attention to the selection of the theoretical framework. We believe that the generalized collective mode (GCM)\index{generalized collective mode (GCM)} approach\cite{GCM1, GCM2, GCM3}, which originates from the method of non-equilibrium statistical operator (NSO)\index{non-equilibrium statistical operator (NSO)} \cite{NSO1, NSO2, NSO3}, is the very tool, being able to explain at least some old puzzles in the fluid dynamics and to foresee or explain the new ones.

The topic we discuss in this Chapter has a long history but many problems are still under discussion. We start in Sec.~2 from short motivation and historical retrospectives, helping the readers to find themselves an answer to the question ``what is a fluid?'' We want to note in advance that the answer is ambiguous: the fluid can be gas-like, liquid-like, or even display the properties of solids, depending on its preparation, external conditions, and the experimental setup.

In Sec.~3 the main ideas of the NSO\index{non-equilibrium statistical operator (NSO)} method are described. The transport equations as well as the equations for the equilibrium time correlation functions (TCFs)\index{time correlation functions (TCFs)} are derived for the case of small deviations from equilibrium. On this basis we consider in more details the equations of generalized hydrodynamics for a multi-component fluid and present expressions for the generalized (dependent on the wavenumber $k$) thermodynamic quantities and the generalized (dependent on the wavenumber $k$ and frequency $\omega$) transport coefficients that can be obtained rigorously within the NSO\index{non-equilibrium statistical operator (NSO)} method. This theoretical framework opens new possibilities for the study of collective dynamics, playing a significant role in fluids, and allows one to use the concept of collective excitations  based on the microscopic  background. In Sec.~4 it is shown how such ideas could be realized within the GCM\index{generalized collective mode (GCM)}.

The applications of the theory, described in Sec.~3 and 4, to several  physical problems are presented in Sec.~5. Here we explore such phenomena as the diffusion in one-component and binary fluids, the viscoelastic properties\index{viscoelastic properties} of fluids under different conditions, the appearance and possibilities of observation of the optical phonon-like excitations in binary and multicomponent dense fluids, which sometimes manifest themselves as a fast sound\index{fast sound}. In addition, the dynamics of molten salts with dominant Coulomb interactions, leading to quite interesting phenomena, are considered. We study also some useful outputs of the theory and discuss the rigorous relations for the transport coefficients. In particular, it is shown that  the so-called ``universal golden rule'' for the partial conductivities in ionic liquids can be simply derived in the generalized form. The nature of the dynamic crossover in supercritical fluids is briefly discussed as well. 

We conclude in Sec.~6 with some final remarks and a brief discussion of the perspectives and some open questions.

\section{What is a fluid?}\label{whatIs}
The Encyclopaedia Britannica defines \textit{a fluid} as 
\begin{quote}
``any liquid or gas or generally any material that cannot sustain a tangential, or shearing, force when at rest and that undergoes a continuous change in shape when subjected to such a stress. This continuous and irrecoverable change of position of one part of the material relative to another part when under shear stress constitutes flow, a characteristic property of fluids.''
\end{quote}
There exists a special branch of physics that explores the flow properties of fluids. \textit{Rheology}\index{rheology} studies the flow of matter, primarily in a liquid state, but also as ``soft solids'' or solids under conditions, in which they react by plastic flow rather than being deformed elastically in response to an applied force. This term was coined by Eugene Bingham in 1920 from the suggestion made by Markus Reiner, inspired by the aphorism attributed to Heraclitus, ``$\Pi\alpha\nu\tau\alpha$ $\rho \epsilon\iota$'' that means ``everything flows'' \cite{Antony-Beris}. 

The ability of a fluid to flow is characterized mostly by its viscosity\index{viscosity}, which is a measure of fluid’s resistance to gradual deformation by shear or tensile stress. A fluid that has no resistance to shear stress is known as an ideal one, and from the physics viewpoint is governed by the dissipationless Euler's equations \cite{EulerEqs}. All real fluids have a non-zero positive viscosity and are described by the Navier–Stokes hydrodynamic equations \cite{deGroot,Sengers}. In such a dissipative system the entropy is being produced, and the motion of the fluid ceases after the external stress is removed.

A fluid with very high viscosity\index{viscosity} may appear as a solid. In this context, one of the most curious experimental setups is worthy to be mentioned. 
\begin{figure}
	\centerline{\includegraphics[width=5cm]{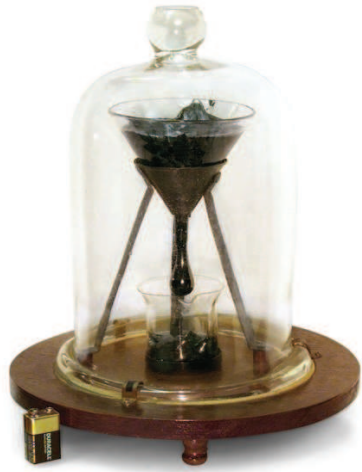}}
	\caption{The University of Queensland pitch drop experiment, demonstrating the visco\-si\-ty of bitumen \cite{pitch-drop}.} \label{fig_p9}
\end{figure}
It is the so-called pitch drop experiment (see Fig.~\ref{fig_p9}), used mainly for educa\-tio\-nal purposes. The long-lasting measurement of the flow of a pitch piece (a highly viscous fluid, a kind of bitumen) was started in 1927 by Prof. Thomas Parnell of the University of Queensland in Brisbane, Australia, to demonstrate to his students that some substances, which look like solids, are, in fact, very-high-viscosity fluids. The eighth drop fell on November 28, 2000, allowing experimentalists to calculate that the pitch viscosity\index{viscosity} is approximately 2.3$\times$10$^{11}$ times greater than that of water. The 9-th drop has separated from the funnel on April 24, 2014. In the same year, another similar experiment, which began in 1914 and predated the Queensland's one by 13 years, was reported in the media. However, the chosen kind of pitch is much more viscous, and this experiment has not yet produced its first drop and is not expected to be completed for at least 1000 years \cite{bergin2014}. 

Elasticity is known as the ability of a body to resist a distorting influence or a deforming force and to return to its original size and shape, when that influence or the force is removed. In a solid, the shear stress is a function of strain (Hooke’s law), but in a fluid, the shear stress is a function of the strain rate (according to Pascal's law). Unlike purely elastic substances, fluids have both elastic and viscous features, and the interplay between them depends on various factors. 
\begin{figure}
	\centerline{\includegraphics[width=5cm,angle=270]{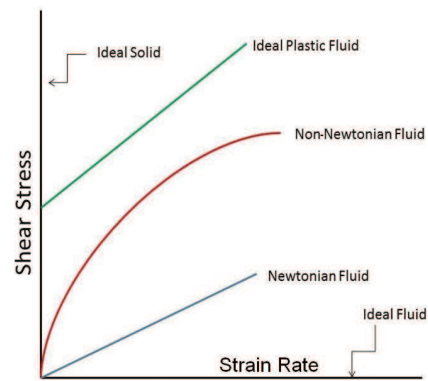}}
	\caption{Shear stress vs. strain rate dependence for various fluids.} \label{fig_p11}
\end{figure}
Depending on the relationship between the shear stress and the rate of strain and its derivatives, fluids are classified as:
\begin{itemize}
	\item 
Newtonian\index{Newtonian fluid} (the stress is directly proportional to the rate of strain);
\item
Non-Newtonian\index{non-Newtonian fluid} (the stress depends on the rate of strain in a more complicated manner, e.g. via its higher powers and derivatives).
\end{itemize}
From the presented in Fig.~\ref{fig_p11} schematic shear stress vs. strain rate diagram for various kinds of fluids it follows that a non-Newtonian fluid\index{non-Newtonian fluid} behaves in a solid-like manner at small strain rates (SR). At the moderate values of SR it loses such features, while at large SR it resembles an ideal fluid (the corresponding curve goes almost parallel to the horizontal axis).

We touch upon this classification of fluids in more detail in Sec.~5, when speaking about the two-variable model for the shear dynamics of a simple liquid. But the question of \textit{how the liquid can manifest its state within the above classification scheme} appears to have a very interesting demonstration in the nature. 

\begin{figure}
	\centerline{\includegraphics[width=11cm]{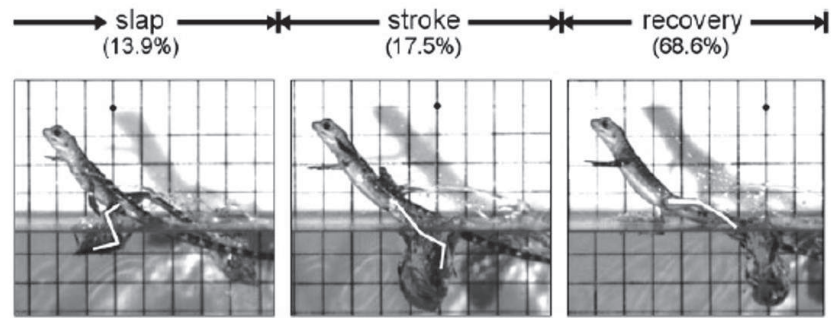}}
	\caption{Running on water: Three dimensional force generating by basilisk lizards \cite{PNAS2004}.} \label{fig_p14}
\end{figure}

In Fig.~\ref{fig_p14}, the snapshots of a basilisk lizards running on water are shown. The Basiliscus (Basilisk) is a genus of large corytophanid lizards, which are endemic to southern Mexico, Central America, and northern South America. They are commonly known as the Jesus Christ lizards, or simply the Jesus lizards, due to their ability to run  on water as a biped for significant distances before sinking. They can do this with the speed of 1.5 meters per second. The running basilisk uses implicitly the solid-like behaviour of liquid when slapping and stroking strongly and frequently on the water surface, whereas it is recovering its position and sliding on the surface due to the viscosity\index{viscosity} of an ordinary (Newtonian) liquid\index{Newtonian fluid}. The juvenile lizards can theoretically generate a maximum total force more than twice their body weight, forcing the ordinary water to behave so extraordinarily (like a non-Newtonian fluid\index{non-Newtonian fluid}). 

Unlike lizards, humans are not able to run on water. This conclusion stems from the famous P.~Kapitsa problem, \textit{to estimate the order of magnitude of the speed with which a man must run so that he would not sink}, posed by him in 1968 in the physics entrance exam. Rough estimations show that the water-runner would need to strike downwards through the water at about 30 m/s, the speed well beyond the human ability. Furthermore, to strike downwards at this speed, he would require an average power output almost 15 times greater than the maximum sustainable output for humans. An interesting study on this subject has been reported in Ref.~\refcite{PLoS2012}.

This problem can be approached in a more quantitative way, using the so-called Deborah number\footnote{The Deborah number was originally proposed by Markus Reiner, a professor at Technion in Israel, who chose the name inspired by a verse in the Bible, stating ``The mountains flowed before the Lord'' in the song by the prophet Deborah in the Book of Judges 5:5.} in order to estimate the ``newtonianity'' of  fluid\index{Newtonian fluid}. The Deborah number, $\mbox{De}=t_r/t_p$, is defined as the ratio of the time it takes for a material to adjust to applied stresses or deformations (relaxation time\index{relaxation time} $t_r$) and the characteristic time scale $t_p$ of an experiment (or of a computer simulation) probing the response of the material. It is often used in rheology\index{rheology} to characterize the fluidity of materials under specific flow conditions \cite{Poole2012}. At lower Deborah numbers, the material behaves in a more fluid-like manner, with an associated Newtonian viscous flow. At higher Deborah numbers, the material behaviour enters the non-Newtonian\index{non-Newtonian fluid} regime, being increasingly dominated by the elasticity and demonstrating the solid-like features. 
%
\begin{figure}
	\centerline{\includegraphics[width=6cm]{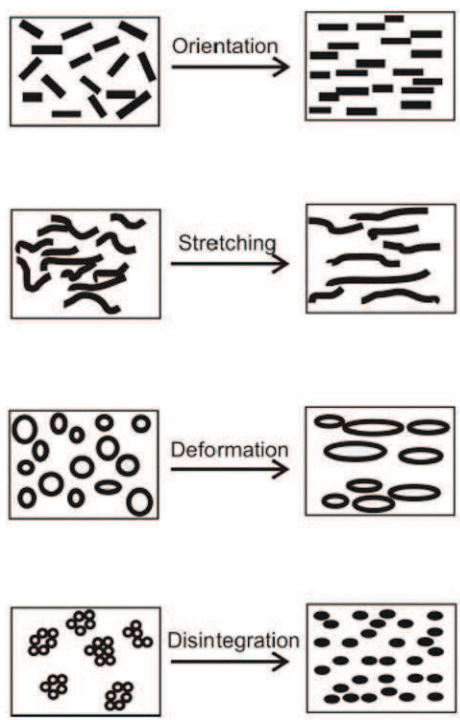}}
	\caption{Viscoelasticity as a reflection of molecular rearrangement.} \label{fig_p15}
\end{figure}

So the Basilisk lizards apparently ``produce'' sufficient Deborah numbers at the surface layers of ordinary water, which in most 
cases is known to be a Newtonian fluid\index{Newtonian fluid} with $\mbox{De}\ll 1$. To observe the non-Newtonian\index{non-Newtonian fluid} behaviour, one has to achieve larger values of the Deborah numbers.  This is possible with increasing the relaxation time\index{relaxation time} $t_r$ and decreasing the contact time $t_p$. Larger relaxation times\index{relaxation time} are typical for fluids that contain particles with a complex macromolecular structure (see, for instance, Fig.~\ref{fig_p15}). It can be expected that the stage of the contact with the liquid surface is very important, because of the opportunity to change the fluid structure locally in the area of contact with the formation of mesoscopic inhomogeneities, observed, in fact, experimentally \cite{PNAS2004}. This leads to an increase of the effective relaxation time\index{relaxation time} $t_r$. Thus, not only the time of contact with the liquid, but also its influence on the liquid local mesostructure in the contact area becomes important. There is a reason to believe that this last factor should be taken into account in order to explain the ability of the Basilisk lizards to run on the surface of water.

Looking at Fig.~\ref{fig_p15}, one can ask another question: ``Is it possible to observe the non-Newtonian\index{non-Newtonian fluid} behaviour in Newtonian fluids\index{Newtonian fluid}? The obvious answer would be: ``Yes, but for the observation times much smaller than the 
relaxation time\index{relaxation time} $t_r$, which is close to the characteristic time of the short-range ordering''. 
\begin{figure}
	\includegraphics[height=2.7cm]{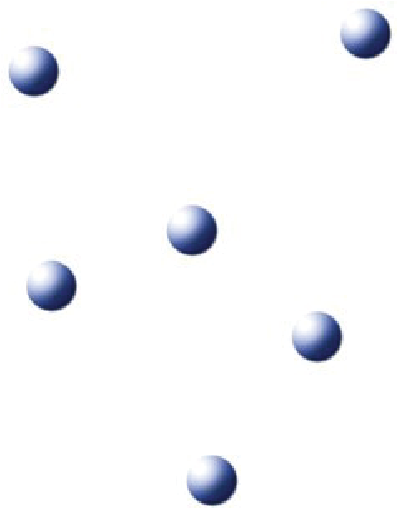}
	\includegraphics[height=2.9cm]{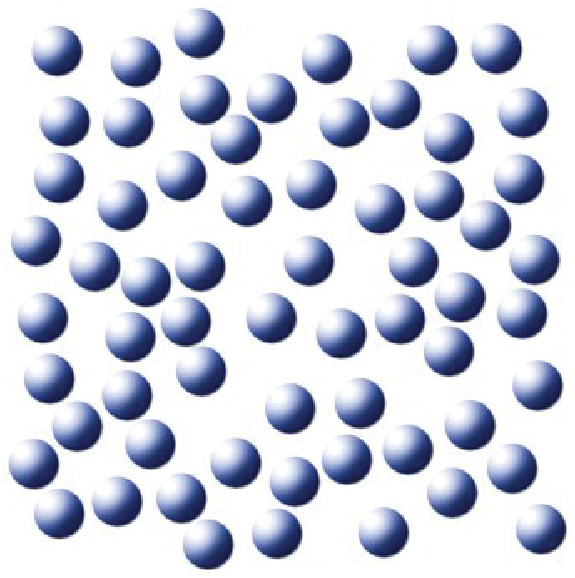}
	\includegraphics[height=2.9cm]{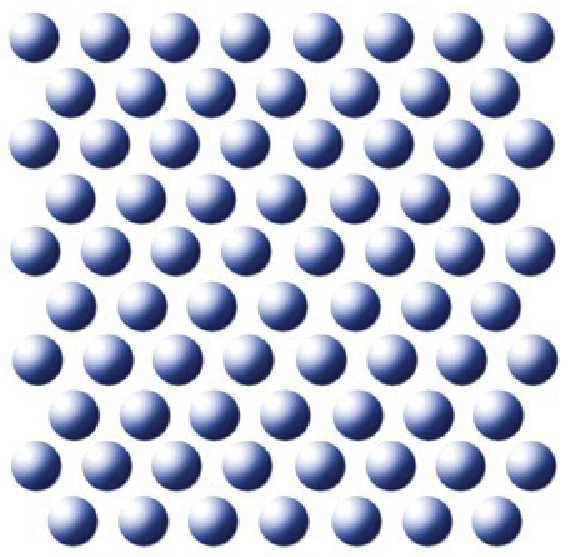}
	\centerline{\includegraphics[width=11.9cm]{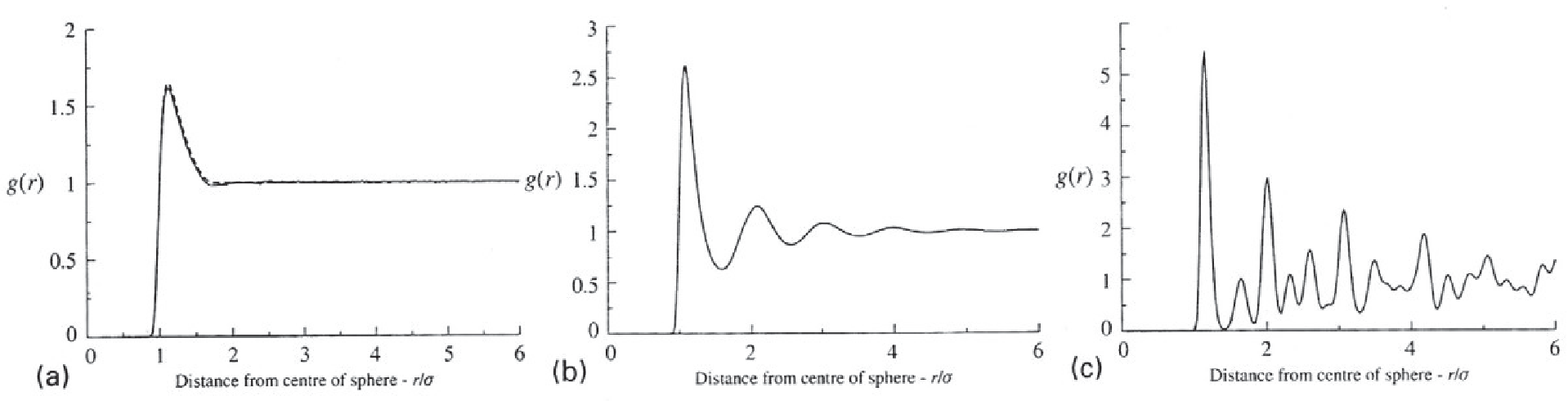}}
	\caption{The structure (upper panel) and the pair correlation functions (lower panel) of a gas (a), liquid or glass (b), and a crystalline solid (c).} \label{fig_p16}
\end{figure}

Liquids are known to posses an intermediate place between  gases and solids with respect to the short-range order.
Unlike dilute gases (see Fig.~\ref{fig_p16}a), whose pair correlation functions have a single maximum at the distances comparable to the particle diameter, fluids have the short-range order, which manifests itself in several maxima of $g(r)$, located in the vicinity of the 1-st, 2-nd and 3-rd coordination spheres (see Fig.~\ref{fig_p16}b). Crystalline solids, on the other hand, are characterized by the long-ranged order, which is just slightly perturbed by the thermal lattice vibrations (phonon modes), leading to the well-pronounced extrema with a rather complicated structure (see Fig.~\ref{fig_p16}c). Thus, the dominant structural ordering processes in a system depend strongly on the phase considered.

It is important to note that  the short-range ordering is typical for dense fluids, just like for liquids. Moreover, the short-range ordering of dynamically arrested particles is one of the characteristic features of glasses. However, this is not the case for dilute gases, although their  symmetry properties are the same as in liquids and glasses. On the other hand, the presence of the short-range order, when a certain particle falls into a fairly stable environment of others, can be reflected in the dynamics as a special form of motion of trapped particles (the cage effect), resembling the phonon motion of particles in crystals. The physics of phonons in crystalline solids with the long-range order is well understood. In liquids, however, the atomic structure is changing with time, and the concept of phonons becomes questionable for long observation times. Therefore, the question arises as to how this process will manifest itself in the collective dynamics of dense fluids, liquids, and glasses. It could give us a new tool to study the phase states, using some dynamic dissimilarities in its collective dynamics.

\begin{figure}
	\centerline{\includegraphics[width=6cm]{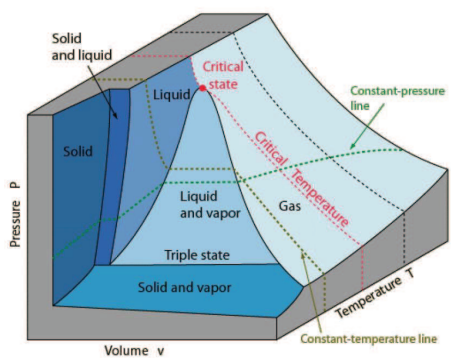}}
	\caption{Typical ($p, V, T$) phase diagram for a simple fluid.}\label{fig_p17}
\end{figure}

A typical phase diagram of a simple fluid is presented in Fig.~\ref{fig_p17}. The gas-liquid critical point is shown by the circle. It is the end point of the pressure–temperature curve that designates conditions under which the li\-quid and the gas can coexist. The first theoretical description of the ``gas--liquid'' phase transition, based upon a hypothesis regarding the form of the $p-V-T$ equation of state, was proposed by Van der Waals in his PhD Thesis \cite{VdW}. Within this simplest theory the critical point can be easily determined.

\begin{figure}
	\centerline{\includegraphics[width=7cm]{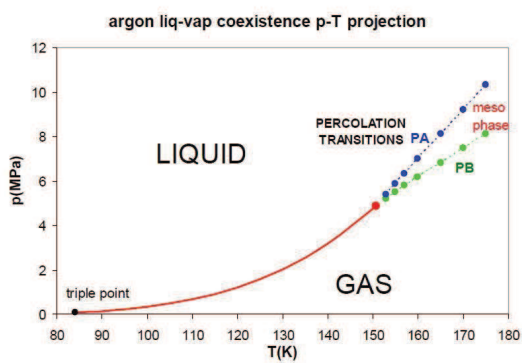}}
	\caption{The liquid-vapor coexistence line for argon from Ref.~\refcite{Woodcock2013}. The available-volume (PA) and bonded-cluster (PB) percolation transitions define the bounds of a supercritical mesophase.}\label{fig_p10}
\end{figure}

Above the critical temperature and the critical pressure the region of a \textit{supercritical fluid} with some very specific features, being interesting for practical applications, is located.  An interesting concept of a supercritical mesophase was proposed recently in Refs.~\refcite{Woodcock2013, Woodcock2014} that, being macroscopically homogeneous but microscopically heterogeneous, exists above the critical point and is bounded by weak higher-order percolation transitions, reflecting some specific structural ordering of particles in a fluid. In the mesophase (see Fig.~\ref{fig_p10}) one can consider the fluid as a microscopically heterogeneous  mixture of gas-like and liquid-like particle clusters that at a certain density suddenly begin to permeate the volume and become macroscopic. Such a new viewpoint provides also the basis for further studies of the dynamical properties and rheological behaviour of supercritical fluids.

\section{Theoretical framework or brief introduction to statistical description}\label{theory}

Basic requirements for the microscopic theory, capable of describing the statistical non-equilibrium thermodynamics and the dynamic properties of fluids, in terms of current knowledge would look as follows:
\begin{itemize}
	\item it has to describe processes on different time and spatial scales;
	\item it has to be flexible and convenient for application as well as for interpretation of the results;
	\item it should be formulated in the computer-adaptable form;
	\item it has to ensure correct thermodynamic and hydrodynamic limits and allow a transparent generalization to a wide range of wave vectors and frequencies. 
	\item one should have control over approximations  (analogue of the sum rules) within the elaborated theory.
		
\end{itemize}

It is clear that the path to creating such a theory was very long and difficult. Although attempts to formulate certain aspects of the kinetic theory of fluids were made already  by Euler, Bernoulli, and Clausius \cite {IndianJournal}, its modern form is based on the contributions of James Clerk Maxwell (1831-1879), Ludwig Boltzmann (1844-1906), and their followers \cite {ISIS,BE}. Since the structure of liquids is much more complicated than that of gases, even numerous generalizations of the Boltzmann kinetic equation failed to describe the liquid dynamics due to the necessity to take the cooperative phenomena into account (the analogue of many-particle collision in the kinetic theory). Less than one century ago, the scientists even questioned the very ability of the statistical physics to describe the liquid state of matter. In 1937, to mark the centenary of Van der Waals, the conference on statistical mechanics was held in Amsterdam, attended by famous scientists such as P.~Debye, G.~Ulenbech, and many others. The session chairman H.~Kramers put the issue ``whether the statistical mechanics can describe the liquid region'' to the vote. An outcome turned out to be a draw: 50 to 50, with P.~Debye voting ``Nay''. Now we know that the correct answer should have been ``Yea''.  But today we may vote ``whether the statistical theory can describe the dynamics of complex fluids''.

Skipping the mid-20th century works by M.~Born, H.~S.~Green and J.~G.~Kirkwood, who had greatly contributed to creation of the kinetic theory of fluids, we have to mention M.~M.~Bogolyubov\footnote{A more ``traditional'' spelling of the name of this great mathematician and physicist of the XX-th century is N.~N.~Bogoliubov, which is a transliteration from the Russian version of the name, Nikolai Nikolajevich Bogoliubov, see Ref.~\refcite{Bogol1946}. However, here we use the transliteration from Ukrainian, Mykola Mykolajovych Bogolyubov. The close ties of M.~M.~Bogolyubov with Ukraine were studied in detail by the authors of this Chapter and the Editor of this volume in Ref.~\refcite{Bogol-and-UA}.} (the main ``co-author'' of the BBGKI hierarchy concept) \cite{Bogol1946}, whose ideas of the abbreviated description of the system dynamics were creatively developed and extended by D.~N.~Zubarev.
\begin{figure}
	\centerline{\includegraphics[width=4.5cm]{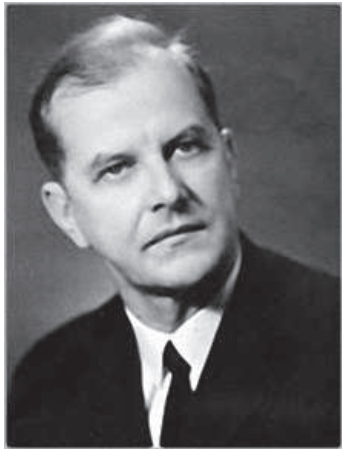}}
	\caption{\textbf{Dmitry Zubarev} (November 27, 1917 -- July 29, 1992),  known for his contributions to statistical mechanics, non-equilibrium thermodynamics, plasma physics, theory of turbulence, and 
		to the development of the two-time Green's function formalism. 
	}\label{fig_p23}
\end{figure}
The main ideas of his method of NSO\index{non-equilibrium statistical operator (NSO)} \cite{NSO1, NSO2, NSO3} are presented below.

Zubarev proposed to start from the time-reversible Liouville equation (LE) for the statistical operator $\rho(\mathbf{x}^N;t)$, depending on the phase space $\mathbf{x}^N = \{\mathbf{p,r} \}^N$ and time $t$, and, using the Bogolyubov's ideas of the weakening of correlations, to average the formal solution of the LE
\begin{equation}\label{rho1}
\rho(\mathbf{x}^N;t)=\exp\{-i \hat L (t-t')\}\rho_q(\mathbf{x}^N;t')
\end{equation} 
(hereafter $\hat L$ means the Liouville operator) with the initial condition $\rho(\mathbf{x}^N;t')=\rho_q(\mathbf{x}^N;t')$ over the time interval $t-t_0$,
\begin{equation}
\rho(\mathbf{x}^N;t)=\frac{1}{t-t_0}\int\limits_{t_0}^t e^{-i \hat L (t-t')}\rho_q(\mathbf{x}^N;t')dt'.
\end{equation}
This interval has to be large enough (more precisely, $t-t_0\to\infty$) to ensure initial states' damping and formation of the dynamical correlations. Using the Abel's theorem \cite{NSO1,NSO2}, one can obtain the LE 
\begin{equation}\label{LE-eps}
\left(\frac{\partial}{\partial t}+i \hat L\right)\rho(\mathbf{x}^N;t)=-\epsilon\{\rho(\mathbf{x}^N;t)-\rho_q(\mathbf{x}^N;t) \}
\end{equation}
with the infinitesimal source $\epsilon$, which violates the time reversibility and tends to zero, $\epsilon\to 0$, after performing the thermodynamic limit $N\to \infty$, $V\to\infty$, $N/V=const$.
The second essential point of the NSO\index{non-equilibrium statistical operator (NSO)} construction, proposed by Zubarev, is a selection of the so-called \textit{quasi-equilibrium (or relevant) statistical operator} $\rho_q(\mathbf{x}^N;t)$ in the generalized Gibbs-like form 
\begin{equation}\label{rho-rel}
\rho_q(\mathbf{x}^N;t)=\exp\left\{-\Phi(t)-\sum\limits_{\alpha}\hat P_{\alpha} F_{\alpha}(t)
\right\}
\end{equation}
in terms of the densities $\hat P_{\alpha}$ of slowly varying dynamical variables of the abbreviated description along with the corresponding conjugate thermodynamic forces $F_{\alpha}(t)$. The first term, $\Phi(t)$, in the exponent of Eq.~(\ref{rho-rel}) denotes the Massier-Planck functional which insures normalisation of the quasi-equilibrium statistical operator to unity.
Since the densities of the dynamic variables describe the local pro\-per\-ties of the system, they depend on the space coordinate $\mathbf{r}$ (or wavevector $\mathbf{k}$ in the Fourier space). Hence, in the above and following expressions an integration over the space coordinate has to be performed explicitly along with the summation over the $\alpha$ index of dynamic variables.

One can obtain the formal solution (\ref{LE-eps}) for the NSO\index{non-equilibrium statistical operator (NSO)} 
\begin{equation}\label{rho2}
\rho(\mathbf{x}^N;t)=\rho_q(\mathbf{x}^N;t)-\!\!\int\limits_{-\infty}^t e^{-\epsilon(t-t')}T(t,t')(1-{\cal P}_q(t'))i\hat L t'\rho_q(\mathbf{x}^N;t') dt'
\end{equation}
as the functional of observables $\mbox{Sp} \{ \hat P_{\alpha}\rho(\mathbf{x}^N;t) \}\equiv\langle\hat P_{\alpha}\rangle^t=\langle\hat P_{\alpha}\rangle^t_q$, where $\mbox{Sp} \{\cdots\}$ means an integration over the phase space for the classical systems (or taking the corresponding trace for the quantum ones), and   $\langle\ldots\rangle_q^t$ means averaging with the quasi-equilibrium statistical operator $\rho_q(\mathbf{x}^N;t)$. In Eq.~(\ref{rho2}) 
\[
T(t,t')=\exp_+\left\{-\int\limits_{t'}^t d\tau(1-{\cal P}_q(\tau))i\hat L
\right\}
\]
denotes the chronologically ordered exponent, constructed on the Kawasaki-Gunton projection operators ${\cal P}_q(t)$, which act in the space of statistical operators \cite{NSO1,MrHach95}.

Using (\ref{rho-rel})-(\ref{rho2}), one can write down the dynamic equations for obser\-vab\-les,
\begin{equation}\label{eqn-Pa}
\frac{\partial}{\partial t}\langle\hat P_{\alpha}\rangle^t=\langle i\hat L P_{\alpha}\rangle^t_q+\sum\limits_{\beta}\int\limits_{-\infty}^t dt' e^{-\epsilon(t-t')}
	\left(\hat I_{\alpha}(t), T(t,t') \hat I_{\beta}(t')
	\right)_q F_{\beta}(t'),
\end{equation}
where
\begin{eqnarray}\label{II-cor}
&&\nonumber
\left(\hat I_{\alpha}(t), T(t,t') \hat I_{\beta}(t')
\right)_q\\ 
&& =\int\limits_0^1 d\tau\langle
\hat I_{\alpha}(t),[\rho_q(\mathbf{x}^N;t')]^{\tau}T(t,t')\hat I_{\beta}(t')[\rho_q(\mathbf{x}^N;t')]^{-\tau}
\rangle_q^t
\end{eqnarray}
denotes the correlation functions, constructed on the generalized fluxes
\begin{equation}
\hat I_{\alpha}(t)=(1-{\cal P}(t))i\hat L\hat P_{\alpha},
\end{equation}
which are defined by the Mori projection operators ${\cal P}(t)$, acting in the space of dynamical variables \cite{MrHach95}.

Assuming the deviation of the generalized thermodynamic forces\index{generalized thermodynamic forces} $\delta F_n(t)=F_n(t)-F_n^0$ from their equilibrium values $F_n^0$ to be small, one can exclude them from expression (\ref{eqn-Pa}) for the observables. In the weakly non-equilibrium case, it is useful to rewrite the transport equation (\ref{eqn-Pa}) in terms of the Fourier transforms $\langle\Delta\hat P_{\alpha}\rangle^{\omega}$ of fluctuations of the dynamical variables $\Delta \hat P_{\alpha}=\hat P_{\alpha}-\langle\hat P_{\alpha}\rangle_0$ around their equilibrium values $\langle\hat P_{\alpha}\rangle_0$ as follows:
\begin{equation}\label{eqn-Paw}
\left\{i\omega {\cal I} -i\Omega_0+\tilde{\varphi}_{\epsilon}(\omega)
\right\}\langle\Delta\hat P_{\alpha}\rangle^{\omega}=0.
\end{equation}
The second term in (\ref{eqn-Paw}) denotes the so-called frequency matrix
\begin{equation}\label{Omega-matrix}
i\Omega=\left(i\hat L \hat P, \Delta \hat P
\right)\left(\Delta\hat P,\Delta\hat P^+
\right)^{-1},
\end{equation}
which describes the non-dissipative processes in the system and is expressed via the corresponding equilibrium correlation functions; ${\cal I}$ means the identity matrix, whose dimension is equal to the number of dynamical variables $\langle\Delta\hat P_{\alpha}\rangle^{\omega}$ considered. The third term in Eqn.~(\ref{eqn-Paw}),
\begin{equation}\label{Phi}
\tilde{\varphi}(\omega)\!=\!\!\left(\!
(1-{\cal P})i\hat L \hat P,\frac{1}{i\omega+\epsilon+(1-{\cal P})i\hat L}(1-{\cal P})i\hat L \hat P^+
\!\right)\!\left(\Delta\hat P,\Delta\hat P^+
\right)^{-1}
\end{equation}
denotes the matrix of memory functions, which determines the dissipation processes in the system.

Since (\ref{eqn-Paw}) represents the system of homogenous linear equations, the determinant of the matrix in the curly brackets has to be equal to zero, defining thereby the spectrum of collective modes. We touch upon this issue in much more detail in Sec.~4, when speaking about the GCM\index{generalized collective mode (GCM)} approach. 
The equation for Laplace transforms $\left(\Delta\hat P,\Delta\hat P^+\right)^z\equiv$ $\left(\Delta\hat P, [z+i\hat L]^{-1}\Delta\hat P^+
\right)$ of the corresponding TCFs\index{time correlation functions (TCFs)} reads in a similar way
\begin{equation}\label{eqn-TCF}
\Big\{i\omega {\cal I} -i\Omega_0+\tilde{\varphi}(z)
\Big\}\left(\Delta\hat P,\Delta\hat P^+\right)^z
=\left(\Delta\hat P,\Delta\hat P^+\right),
\end{equation}
with the r.h.s. of Eq.~(\ref{eqn-TCF}) being the static correlation functions (SCFs)\index{static correlation functions (SCFs)}, constructed on fluctuations of the dynamic variables. It should be stressed that the matrix equation (\ref{eqn-TCF}) for the Laplace transforms of TCFs\index{time correlation functions (TCFs)} is, in fact, an identity. This can be easily proved using the expressions for the frequency matrix (\ref{Omega-matrix}) and the matrix of memory functions (\ref{Phi}). 

Before proceeding to the description of the GCM\index{generalized collective mode (GCM)} approach, it would be instructive to look at the definitions for generalized thermodynamic quantities and generalized transport coefficients\index{generalized transport coefficients} that follow rigorously from the expressions given above. For this purpose, let us consider a $\nu$-component mixture and choose the densities of conserved variables as the basic set of dynamic variables $\hat P_{\mathbf k}=\left\{\{\hat n_{\mathbf k,a}\}, \hat{\mathbf J}_{\mathbf k}, \hat H_{\mathbf k}
\right\}$, where
\begin{equation}\label{nka}
\hat n_{\mathbf k,a}=\sum\limits_{i=1}^{N_a}\exp\{i\mathbf k\mathbf R_i^a\} 
\end{equation}
denotes the number density of particles of the $a$-th species;
\begin{equation}\label{Jk}
\hat{\mathbf{J}}_{\mathbf k}=\sum\limits_a\sum\limits_{i=1}^{N_a}\mathbf p_i^a\exp\{i\mathbf k\mathbf R_i^a
\}
\end{equation}
is the density of the total current;
\begin{equation}\label{Hk}
\hat H_{\mathbf k}=\hat E_{\mathbf k}-\sum\limits_{ab}(\hat E_{\mathbf k},\hat n_{-\mathbf k,a})(\hat n_{\mathbf k},\hat n_{-\mathbf k})^{-1}_{ab}
\hat n_{\mathbf k,b}
\end{equation}
is the enthalpy density, obtained by orthogonalization of the total energy density
\begin{equation}\label{Ek}
\hat E_{\mathbf k}=\sum\limits_a\sum\limits_{i=1}^{N_a} e_i^a\exp\{i\mathbf k\mathbf R_i^a
\}, \quad e_i^a=\frac{(\hat{\mathbf p}_i^a)^2}{2 m_a}+\frac{1}{2}\sum\limits_b\sum\limits_{j=1}^{N_b}V(|\mathbf R_i^a-\mathbf R_j^b|).
\end{equation}
In (\ref{nka})-(\ref{Ek}) the summation over $a, b$ runs from 1 up to the number of species $\nu$; $\mathbf k$ denotes the wavevector; $\mathbf R_i^a$ and  $\mathbf p_i^a$ mean the particle position and momentum, respectively; $m_a$ is the mass of the particle of the $a$-th species, while $V(|\mathbf R_i^a-\mathbf R_j^b|)$ means the pairwise interaction potential.

It was shown in Ref.~\refcite{GCM2} that the generalized ($k$-dependent) thermodynamic quantities can be defined via corresponding SCFs\index{static correlation functions (SCFs)}, constructed on the densities $\hat P_{\mathbf k}$. For instance, the isothermal compressibility $\kappa_T (k)$ can be written as follows
\begin{equation}\label{kT}
\frac{1}{\kappa_T(k)}=N n k_B T\sum\limits_{ab}c_a(\hat n_{\mathbf k}, \hat n_{-\mathbf k})^{-1}_{ab} c_b,
\end{equation}
where $N$ denotes the total particle number, $n=N/V$, $V$ means the system volume, $c_a=N_a/N$ is the concentration of the $a$-th species and $k_B$ is the Bolzmann constant. The specific heat at constant volume $C_V(k)$ is defined by
\begin{equation}\label{CV}
k_B T^2 C_V(k)=(\hat H_{\mathbf k},\hat H_{-\mathbf k}).
\end{equation}
The linear thermal expansion coefficient $\alpha_P(k)$ reads
\begin{equation}\label{alphaP}
\alpha_P(k)=\frac{1}{i k}(i\hat L\hat J_{\mathbf k}^{||},\hat H_{-\mathbf k})\frac{\kappa_T(k)}{k_B T^2 V},
\end{equation}
where $\hat J_{\mathbf k}^{||}$ denotes the longitudinal component of the momentum density. And for the generalized ratio $\gamma(k)=C_P(k)/C_V(k)$ the relation
\begin{equation}\label{gammaK}
\gamma(k)=1+\frac{T V\alpha_P^2(k)}{C_V(k)\kappa_T(k)},
\end{equation}
being well-known from the standard thermodynamics, is satisfied.

In a similar way, the elements of memory functions matrix (\ref{Phi}) can be related to the generalized transport coefficients\index{generalized transport coefficients} \cite{GCM2}, for example:
\begin{equation}\label{PhiNN}
\tilde{\varphi}_{nn}^{ab}(k,z)=k^2 N\sum\limits_c D_{ac}(k,z)(\hat N,\hat N^+)^{-1}_{cb},
\end{equation}
where $D_{ab}(k,z)$ are the generalized mutual diffusion coefficients\index{diffusion coefficients};
\begin{eqnarray}\label{PhiNH}
&&\tilde{\varphi}_{nh}^{a}(k,z)=\frac{k^2}{T C_V(k)}D_T^a(k,z),\\ &&\tilde{\varphi}_{hn}^{a}(k,z)=k^2 V k_B T 
\sum\limits_b D_T^b(k,z)(\hat N,\hat N^+)^{-1}_{ba}, \nonumber
\end{eqnarray}
with $D_T^a(k,z)$ being the generalized thermal diffusion coefficients\index{diffusion coefficients};
\begin{equation}\label{PhiJJ}
\tilde{\varphi}_{JJ}(k,z)=\frac{k^2}{\rho}\eta_l(k,z)
\end{equation}
is connected with the generalized longitudinal viscosity\index{viscosity} $\eta_l(k,z)=4/3 \eta(k,z)+\zeta(k,z)$, where $\eta(k,z)$ and $\zeta(k,z)$ denote, respectively, the gene\-ra\-li\-zed shear and bulk viscosities, and, finally,
\begin{equation}\label{PhiHH}
\tilde{\varphi}_{hh}(k,z)=\frac{k^2 V}{C_V(k)}\lambda(k,z),
\end{equation}
where $\lambda(k,z)$ is the generalized thermal conductivity. There are also the generalized transport coefficients\index{generalized transport coefficients} $\xi(k,z)$ and $\zeta_a(k,z)$,
\begin{eqnarray}\label{PhiCross}
\tilde{\varphi}_{jh}(k,z)=-\frac{i k^2}{T C_V(k)}\xi(k,z),&&
\tilde{\varphi}_{hj}(k,z)=-\frac{i k^2}{\rho}\xi(k,z),\\
\nonumber
\tilde{\varphi}_{nj}^a(k,z)=-\frac{i k^2}{\rho}\zeta_a(k,z),&&
\tilde{\varphi}_{jn}^a(k,z)=-i k^2 T V \sum\limits_b\zeta_b(k,z)(\hat N,\hat N^+)^{1}_{b a},
\end{eqnarray}
which describe the heat--viscosity and diffusion--viscosity dynamic cross-correlations, respectively. Since the corresponding memory functions are constructed on the fluxes of different tensor dimensionality, they contribute in higher order in the wavenumber $k$ and, therefore, can be neglected in the hydrodynamic limit  $k\to 0$.

To summarize this Section, let us emphasize that Eqs.~(\ref{eqn-Paw}) and (\ref{eqn-TCF}) in their present form do not allow one to explore the fluid dynamics, since the memory functions cannot be calculated exactly, and some approximation for $\tilde{\varphi}(\omega)$ is necessary. One of the most efficient ways to proceed is to use the Markovian approximation\index{Markovian approximation} $\tilde{\varphi}(z)\approx\tilde{\varphi}(0)$ for the memory kernels, which from the physical point of view means that all the dissipative processes in the system are  assumed to decay in time very fast. Such an approximation is attractive also because it makes it possible to use computer simulations for the alternative calculations of $\tilde{\varphi}(0)$. However, for many applications  it is important to take into account the $\omega$-dependence of the memory effects. It is possible to do by extending the set of dynamical variables by taking higher derivatives $(i\hat L)^{\alpha}\hat P_{\mathbf k}$, $\alpha>1$, into account, where $\hat P_{\mathbf k}$ denote the densities of the conserved quantities. This idea, which is the basis of the GCM\index{generalized collective mode (GCM)} method, rests upon quite a reasonable assumption that the relaxation times\index{relaxation time} of the kernels, built on the extended set of dynamic variables, are much shorter than those dealt with the basic variables $\hat P_{\mathbf k}$ alone. In such a case the Markovian approximation\index{Markovian approximation} for higher order memory functions is well justified, and it is possible to obtain the closed form of equations (\ref{eqn-TCF}) in terms of the SCFs\index{static correlation functions (SCFs)} and relaxation times\index{relaxation time} for TCFs\index{time correlation functions (TCFs)} of the hydrodynamic origin. Some mathematical aspects of solving Eqs.~(\ref{eqn-TCF}) are considered in the next Section, along with an analysis of the applicability of the GCM\index{generalized collective mode (GCM)} approach from the point of view of the ``sum rules'' requirement.

\section{Generalized collective mode approach}\label{GCM}

Before introducing the basic ideas of the GCM\index{generalized collective mode (GCM)} approach, we would like to discuss briefly the experimental motivation for such a study and the current status of the theory in this domain. To begin with, let us explain the physical meaning of  the  ``collective modes in fluids'' term. The collective modes (or collective excitations) of a certain physical nature, formally decoupled from each other, \footnote{Frequently, such a mode decoupling exists only in a certain domain of the wavenumbers $k$, and there could be an essential overlap of collective excitations at other $k$, which is observed in the shape of the dynamic structure factor or ``momentum-momentum'' spectral function, making it difficult to relate unambiguously a particular mode to certain physical phenomenon. It is an advantage of the GCM\index{generalized collective mode (GCM)} approach that allows to solve such puzzles efficiently in many cases.} 
can be observed, directly or indirectly, in  scattering experiments, measurements of the sound velocity and attenuation, structural relaxation in glasses, peculiarities of the viscoelastic behaviour of fluids, and so on. These modes can be of propagating nature, if they correspond to the sound \cite{GCM3,s-waves}, shear \cite{shear-waves}, heat \cite{h-waves}, charge \cite{JPCM2004}, or magnetic (spin) \cite{spin-waves} waves in fluids. They can be also purely relaxing \cite{GCM3}, when they describe diffusion-like or relaxational processes. Depending on their behaviour at small wavenumbers $k$, they can be divided into the  \textit{hydrodynamic} or \textit{kinetic} subcategories. The hydrodynamic ones are connected with the slowest processes in the hydrodynamic regime, which reflect the dynamics of the conserved quantities and determine the main features of the Navier-Stokes hydrodynamics. Its damping coefficients tend to zero in the  $k \to 0$ limit. Otherwise, the kinetic-like modes tend to non-zero values at $k\to 0$, reflecting the processes of kinetic origin. 

In the middle of the XX century, the so-called \textit{shear waves} were identified in computer simulations. In particular, it has been found that the shape of the transverse spectral function is changed \cite{Boon}, when the wavenumber $k$ exceeds a certain value, typical for each particular liquid. This means that the viscoelastic properties\index{viscoelastic properties} of liquids for smaller spatial scales become similar to those of solids, and liquids behave like elastic bodies. Similar behaviour was observed later for the thermal properties, and this allowed one to identify the so-called \textit{heat waves} \cite{Jose89}. In the second part of the 1980ies,  the so-called \textit{fast sound}\index{fast sound} modes  were found \cite{Bos86} in the scattering experiments, showing the crossover between different types of collective behaviour in dynamics of binary mixtures\index{binary mixtures}. About ten years before that, Hansen and McDonald \cite{Hansen75} carried out a pioneering molecular dynamics (MD)\index{molecular dynamics (MD)} simulation for a model of a molten salt\index{molten salt} and observed the propagating \textit{optic-like modes} in the ``charge-charge'' dynamical structural factors. All these excitations are the examples of propagating modes\index{propagating modes} that could not be described within the standard hydrodynamic theory, so that a lot of different kinds of theories were proposed in order to explain such phenomena \cite{Boon,Hansen,Bal}. However, their main disadvantage was that these theories were formulated mainly for needs of some specific experiments. In addition to the non-hydrodynamic (kinetic) propagating modes\index{propagating modes}, relaxing kinetic-like collective modes were identified in the experiments, that could not be explained within the hydrodynamic theory. For instance, we can mention the \textit{structural relaxation} \cite{glass}, playing a crucial role in the glass-forming liquids, and the \textit{molecular relaxation} \cite{mol-rel}, reflecting the contribution of additional degrees of freedom that are important in complex fluids for certain domains of wavenumbers $k$ and frequencies $\omega$.  

A big challenge for a theory is dealt with the collective dynamics of the multi-component mixtures. In this case, even the problems of hydrodynamic theory have not been completely solved. In particular, no analytical solution is found for the hydrodynamic TCFs\index{time correlation functions (TCFs)} of a mixture with more than three species, indispensable for the correct interpretation of scattering experiments. The question of correct expressions for the hydrodynamic fluxes, that have to be used for calculations of transport coefficients in the multicomponent fluids, is still discussed in literature since this problem is important for many practical applications. Hence we may conclude that there is still a need in the generalized hydrodynamic theories giving us an opportunity to consider all these different phenomena within a unified scheme.

To formalize the main postulates of the GCM\index{generalized collective mode (GCM)} approach, let us consider the extended set of dynamic variables $\hat{\textbf P}^{(s)}=\left\{\hat{\textbf P}_0,\hat{\textbf P}_1,\ldots,\hat{\textbf P}_s\right\}$, $\hat{\textbf P}_s=(i\hat L)^s \hat{\textbf P}_0$,
where $\hat{\textbf P}_0$ denotes the densities of hydrodynamic variables (for instance, the values (\ref{nka})-(\ref{Hk}) in the case of a multi-component fluid). If one passes to the orthogonalized set of the dynamic variables according to the well-known Gramm-Schmidt procedure,
\begin{eqnarray}\label{Ys}
\hat{\mathbf Y}_0=\hat{\mathbf P}_0, \quad\hat{\mathbf Y}_1=(1-{\cal P}_0)\hat{\mathbf P}_1,\ldots,\hat{\mathbf Y}_s=(1-{\cal P}_{s-1})\hat{\mathbf P}_s,
\end{eqnarray}
where the projection operators ${\cal P}_s$ are defined in the standard way,
\begin{eqnarray}\label{deltaPs}
&&{\cal P}_0=\Delta{\cal P}_0=(\ldots,\hat{\mathbf P}_0^+)(\hat{\mathbf P}_0,\hat{\mathbf P}_0^+)^{-1}\hat{\mathbf P}_0,\\
\nonumber
&&{\cal P}_s=\sum\limits_{s'=0}^s\Delta{\cal P}_{s'}, \qquad
\Delta{\cal P}_{s'}=(\ldots,\hat{\mathbf Y}_{s'}^+)(\hat{\mathbf Y}_{s'},\hat{\mathbf Y}_{s'}^+)^{-1}\hat{\mathbf Y}_{s'},\end{eqnarray}
one can generalize the transport equations (\ref{eqn-Paw}) as follows:
 \begin{equation}\label{eqn-Yaw}
 \left\{
 i\omega {\cal I} -i\mathbf{\Omega}^{(s)}+\tilde{
 	\mathbf{\varphi}
}
^{(s)}(\omega)
 \right\}\langle\Delta\hat{\mathbf Y}_s\rangle^{\omega}=0.
\end{equation}
The frequency matrix $i\mathbf{\Omega}^{(s)}$ in Eq.~(\ref{eqn-Yaw}) is of the three-diagonal structure,
\begin{equation}\label{Omegas}
i\mathbf{\Omega}^{(s)}=\left(
\begin{array}{cccccc}
i\mathbf{\Omega}_0&{\cal I}& & & & \\
-\mathbf{\Gamma}_0&i\mathbf{\Omega}_1&{\cal I}& &0\\
 &-\mathbf{\Gamma}_1&i\mathbf{\Omega}_2&{\cal I}& &\\
 & &\ldots&\ldots&\ldots&\\
  &0& &-\mathbf{\Gamma}_{s-2}&i\mathbf{\Omega}_{s-1}&{\cal I}\\
   & & & &-\mathbf{\Gamma}_{s-1}&i\mathbf{\Omega}_s
\end{array}
\right),
\end{equation}
and its components are equal to
\begin{eqnarray}\label{OmegaGamma}
&&i\mathbf{\Omega}_l=i\mathbf{\Omega}_{ll}=(i\hat L\hat{\mathbf Y}_l,\hat{\mathbf Y}_l^+)(\hat{\mathbf Y}_l,\hat{\mathbf Y}_l^+)^{-1},\\
\nonumber
&&i\mathbf{\Omega}_{l-1 l}=(i\hat L\hat{\mathbf Y}_{l-1},\hat{\mathbf Y}_l^+)(\hat{\mathbf Y}_l,\hat{\mathbf Y}_l^+)^{-1}=(\hat{\mathbf Y}_l,\hat{\mathbf Y}_l^+)(\hat{\mathbf Y}_l,\hat{\mathbf Y}_l^+)^{-1}\equiv{\cal I}\\
\nonumber
&&\mathbf{\Gamma}_l=-i\mathbf{\Omega}_{l+1 l}=-(i\hat L\hat{\mathbf Y}_{l+1},\hat{\mathbf Y}_l^+)(\hat{\mathbf Y}_l,\hat{\mathbf Y}_l^+)^{-1}=(\hat{\mathbf Y}_{l+1},\hat{\mathbf Y}_{l+1}^+)(\hat{\mathbf Y}_l,\hat{\mathbf Y}_l^+)^{-1}.
\end{eqnarray}
The matrix of memory functions $\tilde{\varphi}^{(s)}$ in Eq.~(\ref{eqn-Yaw}),
\begin{equation}\label{phis}
\tilde{\varphi}^{(s)}=\left(
\begin{array}{ccccc}
0&0&\ldots0&0\\
0&0&\ldots0&0\\
 & &\ldots& & \\
 0&0&\ldots0&0\\
 0&0&\ldots0&\tilde{\varphi}_s\\
\end{array}
\right),
\end{equation}
has only one non-zero block,
\begin{equation}\label{phis1}
\tilde{\varphi}_s(z)\!=\!\!\left(\!
\hat{\mathbf Y}_{s+1},\frac{1}{z+(1-{\cal P})i\hat L}
\hat{\mathbf Y}_{s+1}^+
\!\right)\!\left(\hat{\mathbf Y}_s,\hat{\mathbf Y}_s^+
\right)^{-1}.
\end{equation}
The system of transport equations (\ref{eqn-Yaw}) can be solved starting from the variables $\langle\Delta\hat{\mathbf Y}_s\rangle^{\omega}$ up to $\langle\Delta\hat{\mathbf Y}_{s-1}\rangle^{\omega}$ (the so-called rolling up procedure). Thus the lower-order memory function $\tilde{\varphi}_{s-1}(z)$ can be related to the higher-order ones $\tilde{\varphi}_{s}(z)$ via the recurrent matrix expression
\begin{equation}\label{recurr}
\tilde{\varphi}_{s-1}(z)=\left[z{\cal I}-i\mathbf{\Omega}_s+\tilde{\varphi}_s(z)
\right]^{-1}\mathbf{\Gamma}_{s-1}.
\end{equation}
Performing the rolling up procedure to the lowest order dynamical variables, one can attribute the obtained expressions at $\langle\Delta\mathbf Y_0\rangle^{\omega}$ to the generalized ($k$ and $\omega$ dependent) transport coefficients \cite{Mry97c}.

The generalization of the dynamic equations for the TCFs\index{time correlation functions (TCFs)} constructed on the orthogonalized basic set (\ref{Ys}) is straightforward (cf. Eq.~(\ref{eqn-TCF})). At this stage, it is well justified to perform the Markovian approximation\index{Markovian approximation} for the memory function  $\tilde{\varphi}_s(z)$. Some general physical reasons to do so have been already formulated at the beginning of this Section. Moreover, there are even stricter ideas based on the so-called ``non-Markovianity spectrum'' elaborated in Refs.~\refcite{Yulmet141,Yulmet142}. We will not go into details of this issue here, but return to it at the analysis of the obtained results.

In the Markovian approximation\index{Markovian approximation}, the equations for TCFs\index{time correlation functions (TCFs)} look as follows:
\begin{eqnarray}\label{eqn-TCF-M}
&&\left[z {\cal I}+\mathbf T^{(s)}(k)
\right]\tilde{\mathbf F}^{(m,s)}(k,z)=\mathbf F^{(s)}(k),\\
\nonumber
 &&\tilde{\mathbf F}^{(m,s)}(k,z)\!=\!\left(\!\Delta \hat{\mathbf Y_{\mathbf k}}^{(s)},[\Delta \hat{\mathbf Y_{-\mathbf k}}^{(s)}]^+\!\right)^z, \,
 \tilde{\mathbf F}^{(s)}(k)\!=\!\left(\!\Delta \hat{\mathbf Y_{\mathbf k}}^{(s)},[\Delta \hat{\mathbf Y_{-\mathbf k}}^{(s)}]^+\!\right),
 \end{eqnarray}
where 
\begin{equation}\label{T-matrix}
\mathbf T^{(s)}( k)=i\mathbf{\Omega}^{(s)}(k)+\tilde{\varphi}^{(s)}(k,0)
\end{equation}
denotes the so-called \textit{generalized hydrodynamic matrix} \cite{GCM1,GCM2}. In Eq.~(\ref{eqn-TCF-M}) we denote the basic set of dynamic variables as
$\hat{\mathbf Y}^{(s)}=\{\hat{\mathbf Y}_{0\mathbf k}, \hat{\mathbf Y}_{1\mathbf k},\ldots, \hat{\mathbf Y}_{s\mathbf k}
\}$, 
whereas the superscript $m$ means the Markovian approximation\index{Markovian approximation}.

To proceed, let us note first that the generalized hydrodynamic matrix (\ref{T-matrix}) can be presented in an alternative way,
\begin{equation}\label{T-matrix1}
\mathbf T^{(s)}(k)=\mathbf F^{(s)}(k)\left[\tilde{\mathbf F}^{(s)}(k,0)
\right]^{-1},
\end{equation}
where the second factor denotes the inverse matrix to that constructed on the corresponding TCFs\index{time correlation functions (TCFs)} at $z=0$. The expression (\ref{T-matrix1}) is more convenient, since $\tilde{\mathbf F}^{(s)}(k,0)$ can be presented in terms of the SCFs\index{static correlation functions (SCFs)} and relaxation times\index{relaxation time} of the hydrodynamic origin \cite{GCM1,GCM2,GCM3}. Note that all the quantities needed for calculations of matrix elements of $\mathbf T^{(s)}( k)$ can be obtained from computer simulations.

The solution of (\ref{eqn-TCF-M}) can be presented in an analytic form via the eigenvalues $z_{\alpha}(k)$ and eigenvectors $||\hat X_{i,\alpha}(k)||$ of the matrix $\mathbf T^{(s)}(k)=||T_{ij}(k)||$,
\begin{equation}\label{eigen}
\sum\limits_{j=1}^M T_{ij}(k)\hat X_{j,\alpha}(k)=z_{\alpha}(k)\hat X_{i,\alpha}(k),
\end{equation}
where the summation runs from 1 up to the modes number $M$. It is obvious that $M$ is equal to the number of all the dynamical variables $\hat{\mathbf Y}^{(s)}$. Moreover, the eigenvalues of $\mathbf T^{(s)}(k)$ can be either real or complex conjugate. In the former case they define the purely relaxing modes\index{relaxing modes}, whereas in the latter case they correspond to the propagating collective excitations.

In terms of the eigenvalues/eigenvectors, the solution of (\ref{eqn-TCF-M})
can be written down as
\begin{equation}\label{sol-z}
\tilde F_{ij}^{(m,s)}(k,z)=\sum\limits_{\alpha=1}^M\frac{G_{\alpha}^{ij}(k)}{z+z_{\alpha}(k)}.
\end{equation}
The numerators in (\ref{sol-z}) are the weighted amplitudes
\begin{equation}\label{Gij}
G_{\alpha}^{ij}(k)=\sum\limits_{l=1}^M\hat X_{i,\alpha}\hat X^{-1}_{\alpha,l}F^{(s)}_{lj}(k),
\end{equation}
which determine the contribution of each particular collective excitation (mode $z_{\alpha}(k)$) to the TCF\index{time correlation functions (TCFs)} $F_{ij}^{(m,s)}(k,z)$.
The expression (\ref{sol-z}) can be converted to time representation,
\begin{equation}\label{sol-i}
 F_{ij}^{(m,s)}(k,t)=\sum\limits_{\alpha=1}^M G_{\alpha}^{ij}(k)\exp\{-z_{\alpha}(k) t\},
\end{equation}
as a sum of weighted exponents.

Some important consequences follow from all the above: 

(i) Due to the linear relations between the initial set of variables $\hat{\mathbf P}_{\alpha\mathbf k}$ and the set of their orthogonalized counterparts $\hat{\mathbf Y}_{\alpha\mathbf k}$, the theory can be easily converted to the form convenient for computer simulations (without projection operators) \cite{GCM3}. 

(ii) The unitary transformation $\hat{\mathbf P}_{\alpha\mathbf k}\rightarrow\hat{\mathbf Y}_{\alpha\mathbf k}$ does not change the collective modes spectrum as it is shown in Ref.~\refcite{Mr1998}. It means that the same modes will be observed in all TCFs\index{time correlation functions (TCFs)} considered, but their magnitudes depend on the values of the corresponding amplitudes in (\ref{sol-i}).

(iii) Within the GCM\index{generalized collective mode (GCM)} approach, one can control  the results for TCFs\index{time correlation functions (TCFs)} that follow from (\ref{sol-i}) via some sum rules for 
the frequency $\langle\omega_l(k)\rangle=\int_{-\infty}^{\infty} d\omega\, \omega^l\tilde F^{(m,s)}(k,\omega)$ and time $\langle\tau_l(k)\rangle=\int_0^{\infty}d\tau\,\tau^l F^{(m,s)}(k,\tau)$ moments of the corresponding TCFs\index{time correlation functions (TCFs)}. It has been proven in Ref.~\refcite{GCM1} that the first ($2s+1$) frequency moments along with the zeroth time moment for each of the TCFs\index{time correlation functions (TCFs)}, calculated in the framework of the GCM\index{generalized collective mode (GCM)} approach, coincide with those associated with the corresponding genuine functions $F_{ij}(k,t)$. Therefore, in order to reproduce  the higher frequency moments explicitly, one has to expand the set of dynamic variables $\hat{\mathbf P}_{\alpha\mathbf k}$ by including the higher order time derivatives.

(iv) At any level of description one can take into account the non-Markovian effects\index{non-Markovian effects} by using the recurrent relations (\ref{recurr}). Such a possibility was considered in our papers \cite{OMT1998,MH1999,CMP2018,JCP2018}, using the concept of converging of the relaxation times\index{relaxation time} in the memory functions of the highest order. Formally this is taken into account by approximation $\tilde{\varphi}_{s-1}(z)\approx\tilde{\varphi}_s(z)$ in the recurrent relation (\ref{recurr})).  It is shown that such a modification of the GCM\index{generalized collective mode (GCM)} approach allows to reproduce correctly one extra frequency moment even without expanding the basic set of variables. This modification of the GCM\index{generalized collective mode (GCM)} approach is closely related to the above mentioned spectrum of non-Markovianity\index{non-Markovian effects} \cite{Yulmet141,Yulmet142}. 

In the next Section various dynamic phenomena, arising in the fluids, are analysed within the framework of GCM.

\section{Some problems related to the fluid dynamics}

\subsection{Single particle diffusion}

\begin{figure}[h]
	\centerline{\includegraphics[width=7cm]{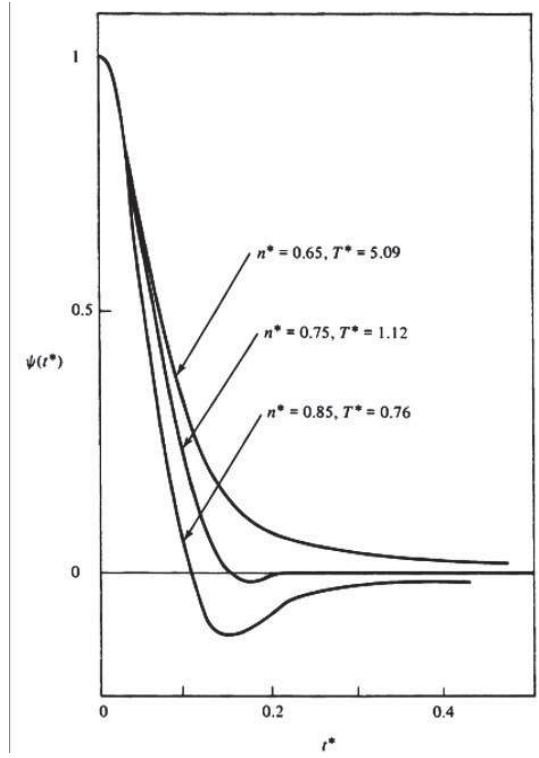}}
	\caption{
		Velocity autocorrelation function for the 6:12 Lennard-Jones fluid at various densities and temperatures, as obtained from computer simulation experiment  \cite{Levesque-Verlet}. Reduced units are $n^*=n\sigma^3_{LJ}$, $T^*=k_B T/\varepsilon_{LJ}$, $t^*=t(\varepsilon_{LJ}/M\sigma_{LJ}^2)^{1/2}$.
	} \label{fig_p33}
\end{figure}

Let us start our consideration of some puzzling phenomena in the fluid dynamics from the simplest example of TCFs\index{time correlation functions (TCFs)}: the single-particle velo\-city autocorrelation function (VAF), calculated usually in MD\index{molecular dynamics (MD)} simulations. It is interesting that this function behaves quite differently in low- and high-density systems, where the so-called \textit{cage effect} can be observed. The cage effect reflects a single particle trapping by the local configuration of its neighbouring molecules, which usually manifests itself as the negative domain (see Fig.~\ref{fig_p33}) of the VAF
\begin{equation}\label{VAF-def}
\psi(t)=\frac{1}{3}\langle
\mathbf v_i(t)\cdot \mathbf v_i(0)
\rangle,
\end{equation}
where $\mathbf v_i(t)$ denotes the $i$-th particle velocity at the time instant $t$, and $\langle\ldots\rangle$ means the thermal averaging.
Since the self-diffusion coefficient\index{diffusion coefficients} $D$ and the VAF (\ref{VAF-def}) are connected by a simple relation $D=\int_0^{\infty}\psi(t)dt$, it means that at high density some particles are trapped by the surrounding molecules, and its motion to the neighbouring positions of local equilibrium is hampered to a great extent.

Within the GCM\index{generalized collective mode (GCM)} approach such a behaviour of the VAFs can be well described \cite{BMT2003} even at the three-variable basis $\hat{\mathbf P}^{(3)}=\left\{ n(\mathbf k,t),\dot n(\mathbf k,t),\ddot n(\mathbf k,t) \right\}$, where the dynamical variable $n(\mathbf k,t)=\exp(i\mathbf k\mathbf r_j(t))$ defines the self-intermediate scattering function $F(k,t)=\langle n(\mathbf k,t) n(-\mathbf k,0) \rangle$.
The modes' analysis shows \cite{BMT2003} that there are two propagating kinetic modes if $\displaystyle m D\Omega_E <2k_B T$,
where $\Omega_E$ denotes the so-called Einstein frequency defined by the ``force-force'' SCF\index{static correlation functions (SCFs)}. Like with other kinetic modes, the ampli\-tu\-des of these single-particle excitations  vanish in the hydrodynamic limit $k\to 0$, thereby yielding the well known single exponential dependence $F(k,t)=\exp(-D k^2 t)$. 

Using the relation \[\psi(t)=\lim\limits_{k\to 0} \left\{-\frac{1}{k^2}\frac{d^2}{dt^2} F(k,t)\right\}\]
 between the VAF and self-intermediate scattering function, we can express $\psi(t)$ as a weighted sum of two oscillating terms \cite{BMT2003} with damping $\tilde{\sigma}$ and renormalized frequency $\tilde{\omega}$,
\begin{eqnarray}\label{sigma-omega}
&&\tilde{\sigma}=\frac{\tau_{\mathrm{dif}}}{2\tau^2_{\mathrm{vib}}},\quad \tilde{\omega}=\Omega_E\left[1-\left(\frac{\tau_{\mathrm{dif}}}{2\tau_{\mathrm{vib}}}
\right)^2
\right]^{1/2},
\\
\nonumber
&&\tau_{\mathrm{dif}}=\frac{m D}{k_B T},\quad \tau_{\mathrm{vib}}=\frac{1}{\Omega_E},
\end{eqnarray}
where two specific times $\tau_{\mathrm{dif}}$ and $\tau_{\mathrm{vib}}$ characterize, respectively, the hydro\-dyna\-mic processes, connected with the self-diffusion, and the processes of vibrational nature, strongly dependent on an arrangement of the tagged particle. It is obvious that the condition $\tau_{\mathrm{dif}}\ll\tau_{\mathrm{vib}}$ presumes the oscillating behaviour with a well-defined minimum of the VAF; conversely, if the times are close, the cage effect cannot be observed. So, one can conclude that even in the single-particle dynamics of fluids the collective properties in local arrangement play an important role. 
\begin{figure*}[htb]
	\centerline{\includegraphics[height=0.19\textheight,angle=0]{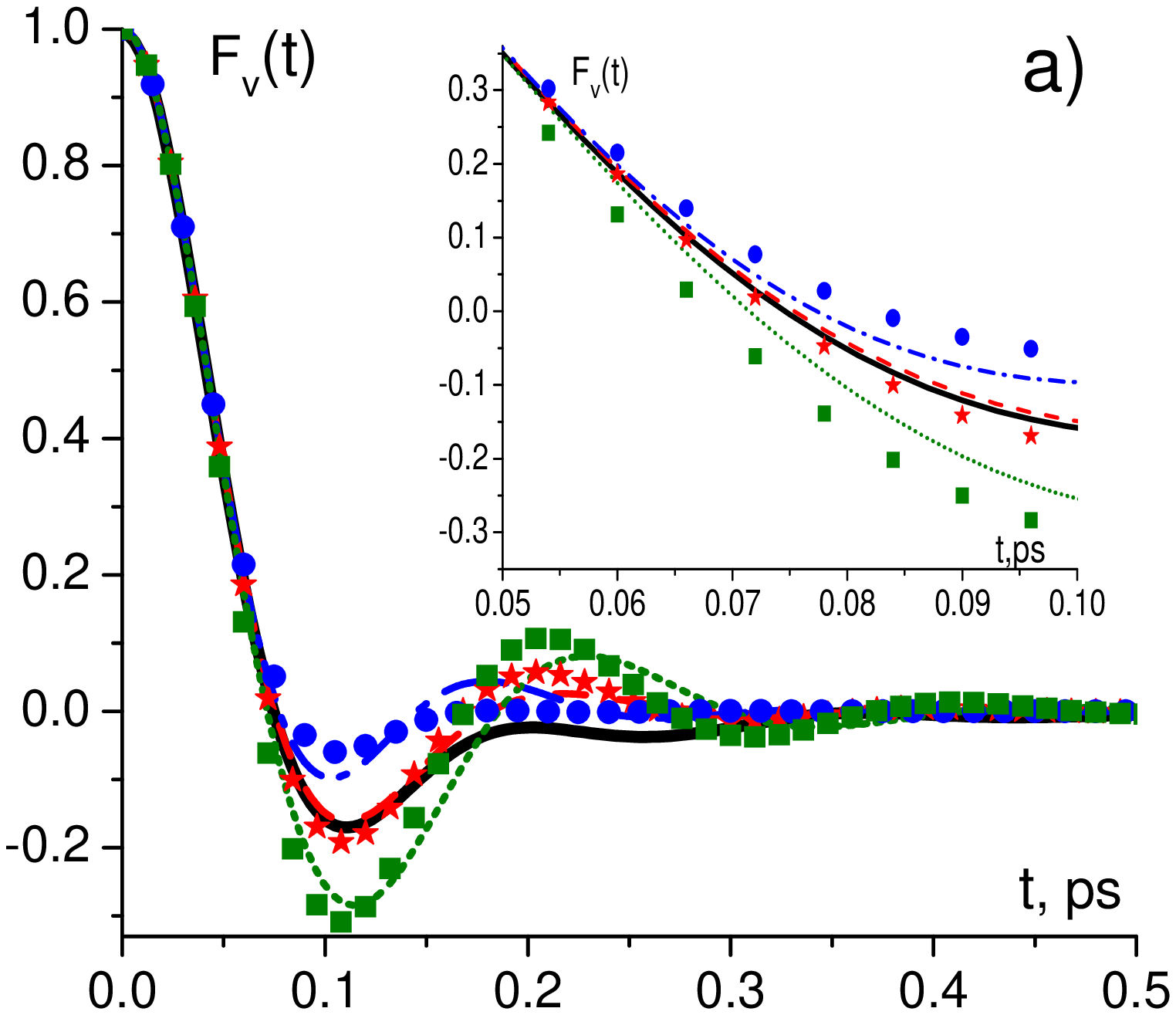}
		\includegraphics[height=0.19\textheight,angle=0]{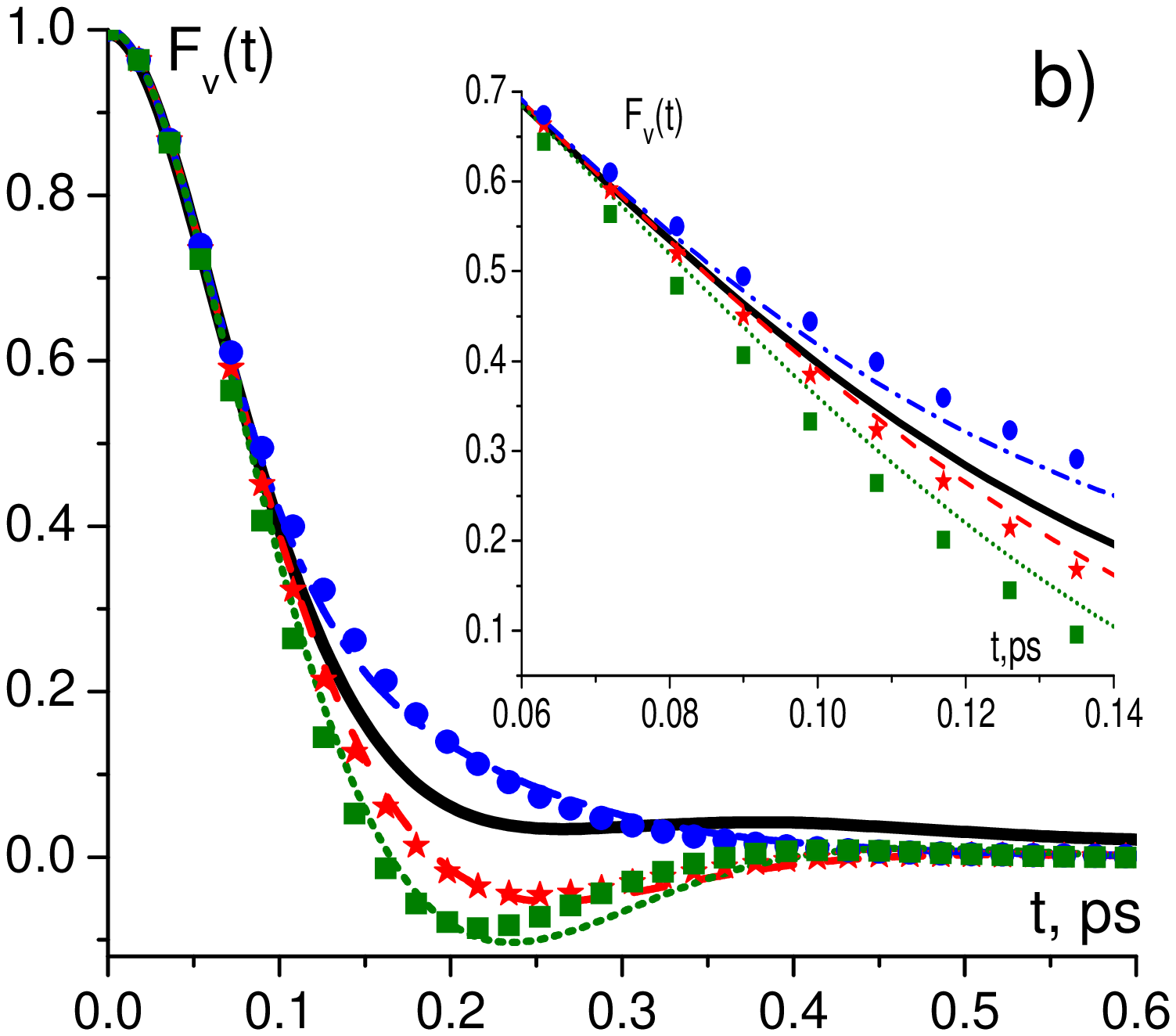}
		\includegraphics[height=0.19\textheight,angle=0]{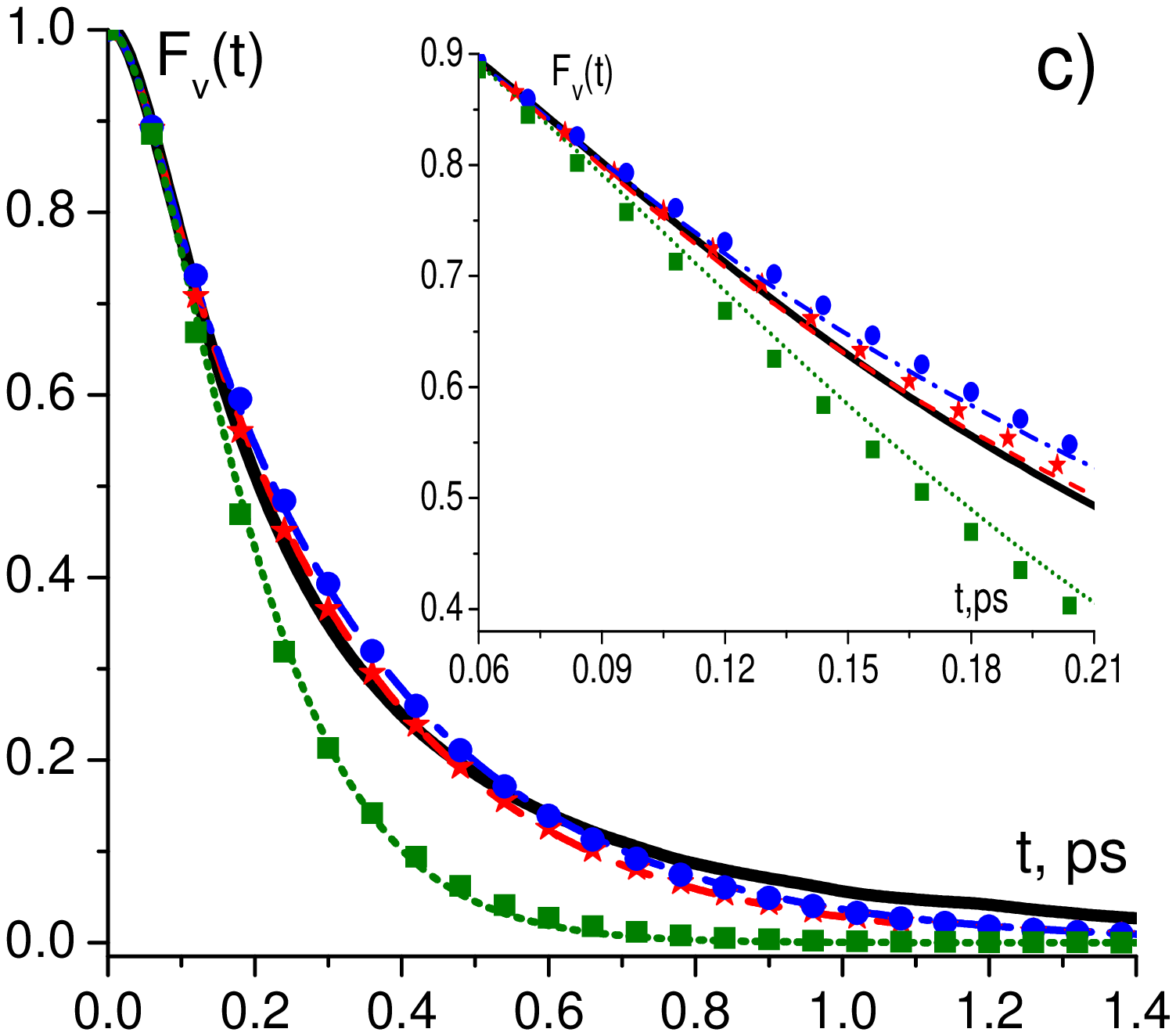}}
	\caption{VAFs of the liquid Ne at the densities $\rho=$2190 kg/m$^3$ (a), 1400 kg/m$^3$ (b)
		and 691 kg/m$^3$ (c). The solid lines correspond to the MD data. Other lines correspond to the VAFs, calculated with the help of (\ref{recurr}) on the 3-rd (dotted line), 4-th (dashed-dotted line) and 5-th (dashed line) hierarchy levels. The symbols correspond to the VAFs, obtained within Markovian approximation for memory functions of the 3-rd  (squares), 4-th (bullets) and 5-th (stars) levels. In the insets: time behavior of the corresponding VAFs at small $t$. The results are obtained in Ref.~\refcite{JCP2018}.}
	\label{figs-JCP}
\end{figure*}

At the quantitative level, these results can be significantly improved in comparison with those obtained in MD\index{molecular dynamics (MD)} simulations, by taking into account the non-Markovian effects\index{non-Markovian effects}. For this purpose one can put $\tilde{\varphi}_{s}(z)\approx\tilde{\varphi}_{s+1}(z)$ in the recurrent relations (\ref{recurr}). The obtained results (see Ref.~\refcite{JCP2018}) are presented in Fig.~\ref{figs-JCP}.
One can draw several conclusions from it. First of all, a convergence of all results to the MD\index{molecular dynamics (MD)} data (denoted by the solid master curve) with increasing of the approximation order is clearly noticeable. Both lines (non-Markovian effects\index{non-Markovian effects}) and symbols (Markovian approximation\index{Markovian approximation}) follow the solid black line the better, the higher order hierarchy $s$ is taken into
account. It is quite expected, since by increasing $s$ we ensure more frequency moments of the VAFs to be satisfied. 

Moreover, the modified version of the GCM\index{generalized collective mode (GCM)} approach with the effective summation of the continued fraction, based on the assumption  $\tilde{\varphi}_{s}(z)\approx\tilde{\varphi}_{s+1}(z)$, allows one to look from a new viewpoint at the formation of the long-time
tails of VAFs $\sim t^{-3/2}$. Recently, a micro\-scopic interpretation of this phenomenon was suggested from analysis of the memory
kernels \cite{27inJCP}. It was shown that the hydrodynamic added mass, defined via the me\-mo\-ry kernel, is negative, and the
backflow of neighbours tended to drag the particle in the direction of its initial velocity, i.e., contributed negatively to the
friction. The approach proposed in \cite{OMT1998,MH1999} and developed in \cite{CMP2018,JCP2018} allows to make one further step and to evaluate the transition time to the long-time tails appearance.\footnote{The effective time between particle collisions can
	be related to the instant, when the ``force–force'' autocorrelation
	function changes its sign. Besides, the well depth in the above mentioned function strongly decreases with decreasing density, like it also happens in the VAFs. This phenomenon can be explained in the following
	way: at low densities, the particle changes its direction
	of motion due to the multiple low-angle scattering, whereas
	at high densities, the ``head-on'' collisions dominate, yielding the profound minimum in ``force–force'' autocorrelation
	function.}
This is another manifestation of a close relationship between the single particle dynamics and the collective behavior of its surrounding.

This issue has been elaborated in much more detail in Ref.~\refcite{FMPF}, where the transition regimes and the shapes of the VAFs are studied for the binary mixture\index{binary mixtures} of particles with large mass asymmetry. The mixtures with the light particles concentration $x=0.2$ and various ratios $\mu$ of heavy to light particle masses were considered.
\begin{figure}[h]
	\centerline{\includegraphics[width=8cm]{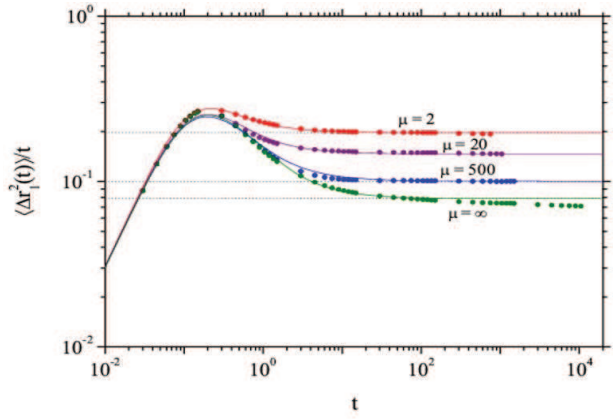}}
	\caption{
		Points: MD data of the MSD of the light
		species divided by time for $x=0.2$ and mass ratios $\mu=2$ (red),
		$\mu=20$ (purple), $\mu=500$ (blue), and $\mu=\infty$ (green). Solid lines: mean field
		results, fitted for the mean free path length $l$, which are 0.15,
		0.15, 0.155, and 0.164, respectively. Dashed lines: diffusion coefficients\index{diffusion coefficients}
		obtained from the Green-Kubo relation multiplied by 6. The dimensionless
		density and temperature equal $\varrho=0.9$ and $T=1$, respectively \cite{FMPF}.
	} \label{fig_p34-1}
\end{figure}
In Fig.~\ref{fig_p34-1}, the mean square displacement (MSD) to time ratio is plotted as a function of $t$. In general, one can distinguish three regimes of the MSD: the quadratic regime for small times, where the particles move ballistically at constant velocity, the linear regime with the usual diffusion coefficient\index{diffusion coefficients} $D_{\alpha}=\lim\limits_{t\to\infty}\langle\Delta r_{\alpha}^2\rangle/6 t$ for large times, and the intermediate region of anomalous diffusion.
Such diffusion is often explained by the cage effect, when particles are trapped inside the cage formed by their surrounding neighbours for some time before they can escape it and diffuse in the usual way. As seen from Fig.~\ref{fig_p34-1}, the region of  the anomalous diffusion increases with $\mu$. This supports the idea \cite{39inFMPF} that the trajectories of the light
particles for large enough $\mu$ change from relatively smooth (Gaussian-like process) to intermittent ones with a large
amplitude of displacements (highly non-Fickian process or activated hopping), as it is shown in Fig.~\ref{fig_p34-2}. 
\begin{figure}
	\centerline{\includegraphics[height=4 cm]{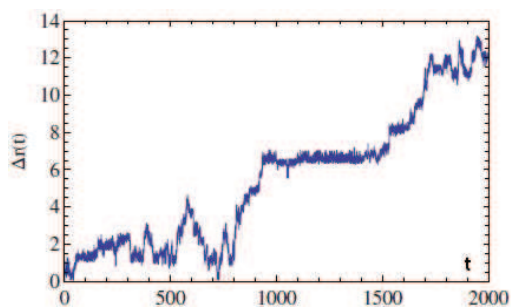}}
	\caption{
		Displacement of the light particle in the mixture with $\mu=\infty$. The other system parameters are the same as in Fig.~\ref{fig_p34-1}\cite{FMPF}.
	} \label{fig_p34-2}
\end{figure}
It is clearly seen that the particle is repeatedly trapped, oscillating around a certain point with an amplitude
smaller than the particle size, before it hops again to some other trap. Hence its displacement demonstrates well the discussed above intermittent process, which is known to be a specific feature of the self-diffusion \cite{FMPF}.

It is also interesting to look at the time dependence of the VAFs of light and heavy particles.
\begin{figure}
	\includegraphics[width=5.5cm]{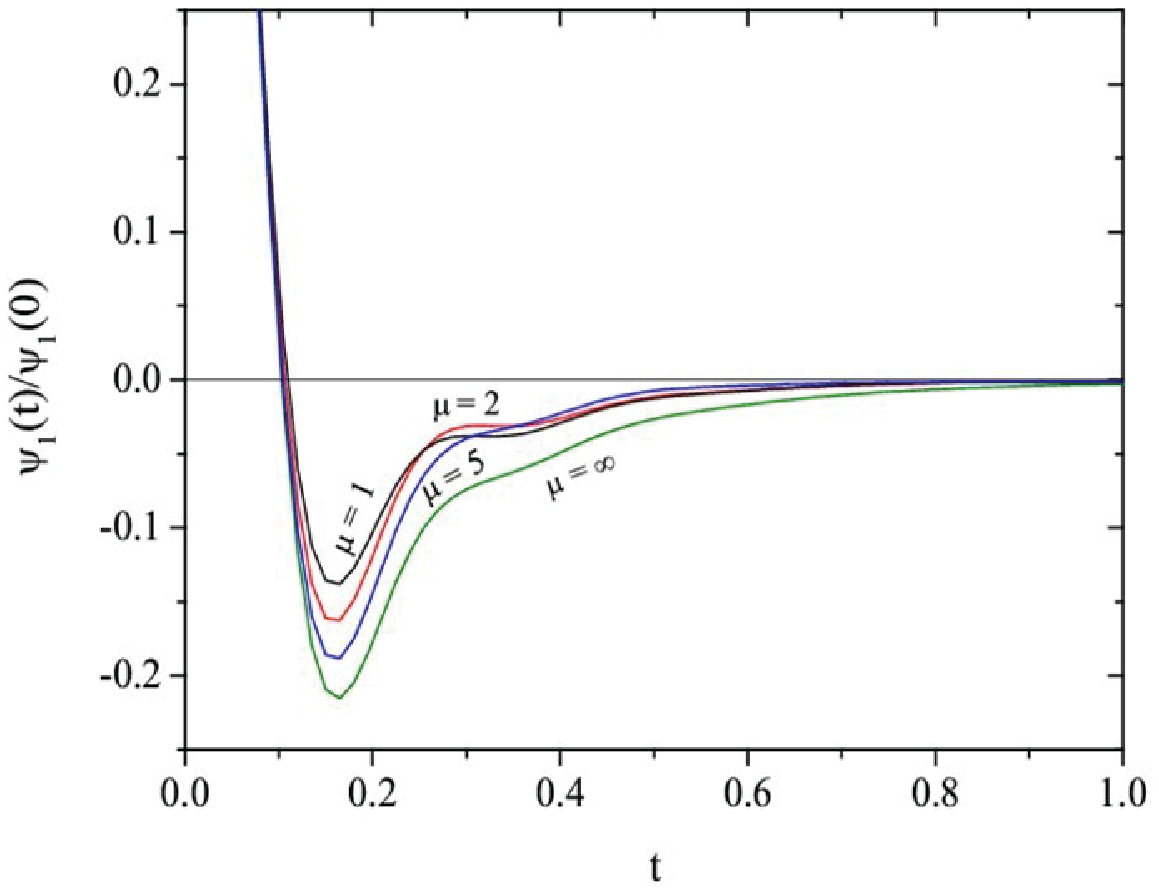}
	\quad
	\includegraphics[width=5.5cm]{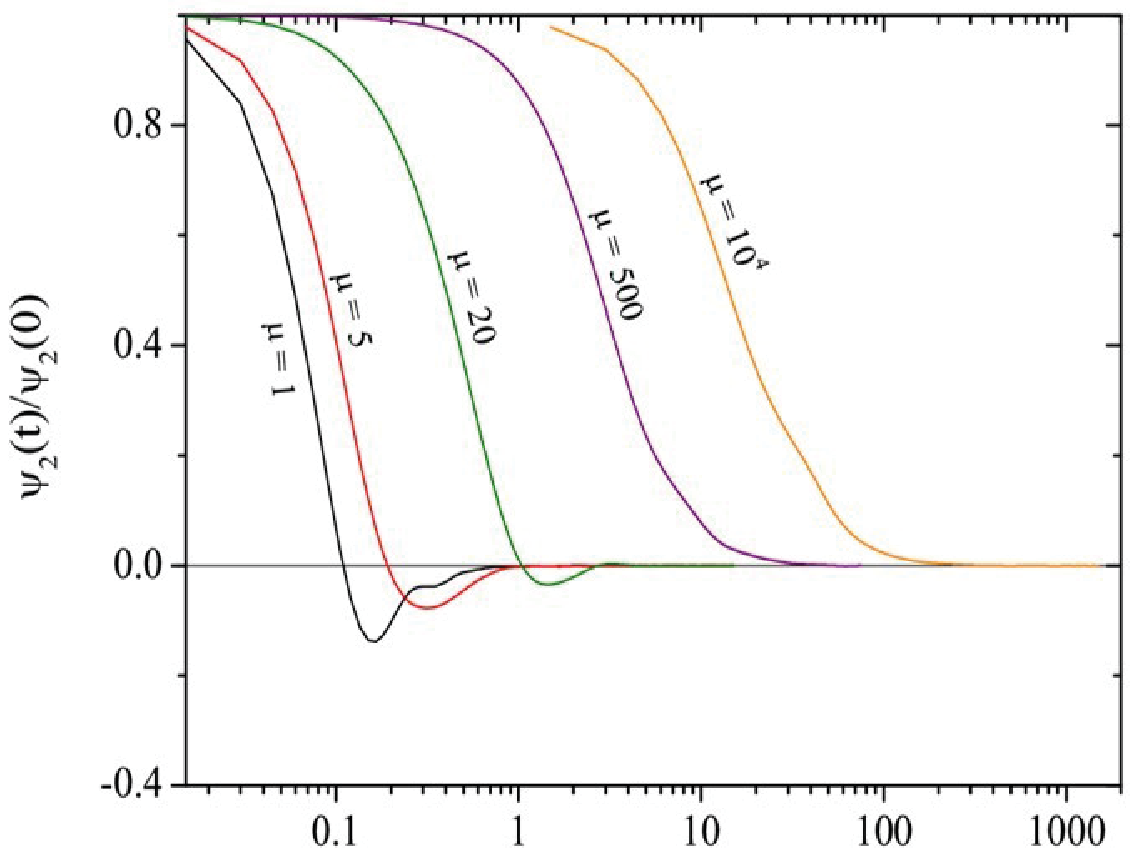}
	\caption{
		Normalized VAFs of the light (left panel) and the heavy particles (right panel)
		at different mass ratios $\mu$ \cite{FMPF}.
	}%
	\label{fig_p35}
\end{figure}
It is seen in Fig.~\ref{fig_p35} that the VAF of the light particles shows a distinct negative minimum, which becomes more pronounced if $\mu$ increases. On the other hand, in the VAF of the heavy particles the minimum vanishes with increasing mass ratio $\mu$. The position of the minimum $t_{min}$ offers a way to estimate the typical size of a cage. With the mean thermal velocity of $v_0=3$ at $T=1$ the light particles on average travel a distance $v_0 t_{min}=0.26$ until they are reflected by the surrounding cage. Together with the particle radius of 0.5 this results in the cage diameter of about 1.5. 

It has also to be stressed that in Ref.~\refcite{FMPF} the extensive numerical simulations have been performed
in order to verify whether the traps are indeed minima of the
potential energy landscape created by the fixed particles. Thus this study gives in fact an example
that a rather stable ``solvent cage'' can be formed in the
mixtures just because of a strong mass asymmetry effect.

\subsection{Viscoelastic properties\index{viscoelastic properties} of fluids}
The GCM\index{generalized collective mode (GCM)} framework, which manifested its efficiency for investigation of the single particle excitations in the VAFs of fluids, can be just as effectively applied for collective phenomena. We start from the simplest case of transverse dynamics in simple liquids. 
It is well known from numerous MD\index{molecular dynamics (MD)} simulations that a crossover from the viscous to elastic behaviour is observed \cite{Mry97a,JFS1998,PhylMag2020} in the shape of the transverse spectral function that can be treated as a signal of the shear waves appearance.

The simplest dynamic model that allows us to understand such a type of crossover can be obtained if we consider the set of two dynamic variables, one of which is dealt with the transverse current density $\hat{\mathbf J}^{\perp}_{\mathbf k}$, and the other one is related to the first  derivative of this variable $i\hat L\hat{\mathbf J}^{\perp}_{\mathbf k}$. In such a case, the equations of macrodynamics can be written down as follows:
\begin{eqnarray}\label{eqn-transverse}
&&
i\omega\langle \hat{\mathbf J}^{\perp}_{\mathbf k}\rangle^{\omega}-\langle i\hat L\hat{\mathbf J}^{\perp}_{\mathbf k}\rangle^{\omega}=0,\\
\nonumber
&&
\left[i\omega+\frac{1}{\tau(k)}\right]
\langle i\hat L\hat{\mathbf J}^{\perp}_{\mathbf k}\rangle^{\omega}+\Gamma_0(k)\langle \hat{\mathbf J}^{\perp}_{\mathbf k}\rangle^{\omega}=0,
\end{eqnarray}
where the inverse relaxation time\index{relaxation time} $1/\tau(k)|_{k\to 0}\simeq G/\eta$ is related to the me\-mo\-ry kernel, constructed on the variables $i\hat L\hat{\mathbf J}^{\perp}_{\mathbf k}$, whereas $\displaystyle\Gamma_0(k)=k^2G(k)/\varrho$ is dealt with the generalized rigidity modulus $G(k)$, defined by the ``stress-stress'' SCF\index{static correlation functions (SCFs)}, 
$$G(k)=\displaystyle\frac{n}{k_B T}\left\langle\hat\sigma^{\perp}_{\mathbf k} \hat\sigma^{\perp}_{-\mathbf k}\right\rangle,$$
$\varrho=n m$ and $\eta$ are the mass density and the shear viscosity\index{viscosity}, respectively. 

A similar chain of equations can be written down for the TCFs\index{time correlation functions (TCFs)} $F_{J^{\perp}J^{\perp}}(k,t)$ built on the transverse components of the current density, and the solution for its normalized value in the time representation looks as follows:
\begin{eqnarray}\label{eqn-TCF-tr}
\frac{F_{J^{\perp}J^{\perp}}(k,t)}{F_{J^{\perp}J^{\perp}}(k)}=-\frac{z_-(k)}{z_+(k)-z_-(k)}e^{-z_+(k)t}+\frac{z_+(k)}{z_+(k)-z_-(k)}e^{-z_-(k)t},
\end{eqnarray}
where the collective modes $z_{\pm}(k)$ are defined by the expression 
\begin{equation}\label{ztrans}
z_{\pm}(k)=\frac{1}{2\tau(k)}\pm\left[\frac{1}{4\tau^2(k)}-\frac{k^2}{\varrho}G(k)
\right]^{1/2}.
\end{equation}
It can be easily verified that the first three sum rules for genuine TCF\index{time correlation functions (TCFs)}  $F_{J^{\perp}J^{\perp}}(k,t)$ are satisfied by the expression (\ref{ztrans}). Moreover, the hydrodynamic correlation time for $F_{J^{\perp}J^{\perp}}(k,t)$ for this simplest non-trivial model (\ref{eqn-transverse}) can be calculated within the GCM\index{generalized collective mode (GCM)} approach. 

In the hydrodynamic limit $k \to 0$, the first term in Eq.~(\ref{eqn-TCF-tr}), being proportional to $k^2$, tends to zero, and the ``current-current'' TCF\index{time correlation functions (TCFs)} is reduced to its well-known single exponent form of a purely diffusive nature,
\begin{equation}\label{TCF-JJ}
\frac{F_{J^{\perp}J^{\perp}}(k,t)}{F_{J^{\perp}J^{\perp}}(k)}\simeq\exp\left(-\frac{\eta}{\varrho}k^2 t\right).
\end{equation}
Contrary, at $k>k_H=\sqrt{\varrho G}/2\eta$ there exist two propagating modes\index{propagating modes}, as it is clearly seen from (\ref{ztrans}). Therefore, the kinetic propagating modes related to the shear waves are detectable \cite{Mry97a,JFS1998} in the transverse ``current-current'' TFC, starting from some threshold wavenumber $k > \sqrt{2} k_H$. 
\begin{figure}
	\centerline{\includegraphics[width=10cm]{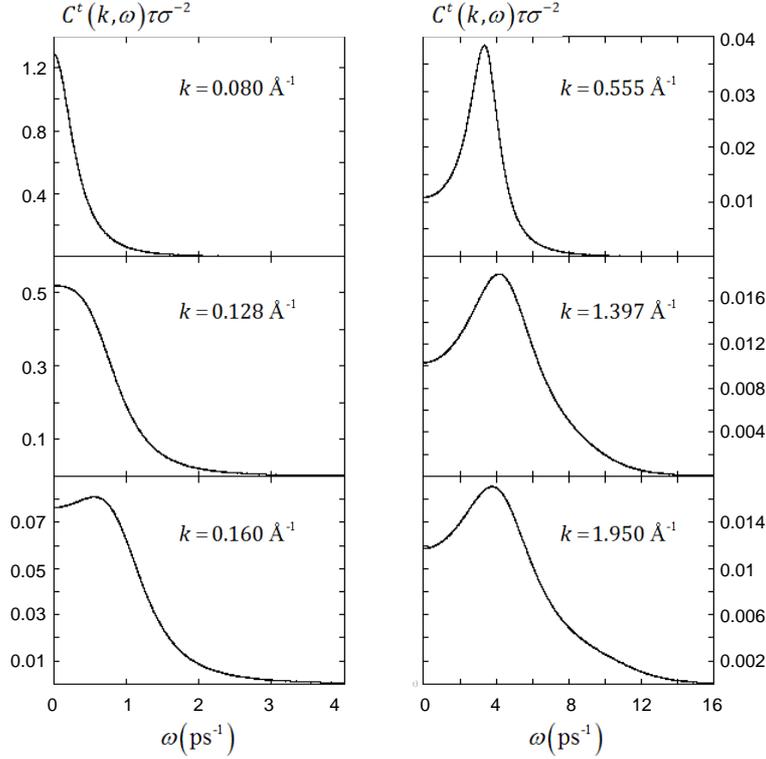}}
	\caption{
		The Fourier-transforms of transverse ``current-current'' TCFs\index{time correlation functions (TCFs)} of liquid Cs at $\varrho=1832$ kg/m$^3$ and  $T=308$ K for six values of $k$. The time and length scale units are $\tau_{\sigma}=4.494$ ps and $\sigma=k_{min}^{-1}=6.239\AA$, respectively.
	} \label{fig_p36}
\end{figure}

The above mentioned phenomenon is illustrated in Fig.~\ref{fig_p36}, where the Fourier spectra of the transverse ``current-current'' TCF\index{time correlation functions (TCFs)} of liquid Cs, obtained in Ref.~\refcite{JFS1998} within the GCM\index{generalized collective mode (GCM)} approach for the dynamic model with four variables, $M=4$, are shown. For $k<\sqrt{2}k_H$, the functions $C^t(k,\omega)$ are single Lorentzians (cf. Eq.~(\ref{TCF-JJ})) centred at the origin.\footnote{At higher modes approximation, the value $k_H$ cannot be obtained analytically, and only numerical calculation of the $T(k)$ matrix spectrum is possible.} For $k>\sqrt{2}k_H$, the functions $C^t(k,\omega)$ have a peak centred at $\omega > 0$. The strong contribution of the pronounced non-central Lorentzian is the consequence of propagating modes\index{propagating modes}
with small damping associated with the real part of the relevant eigenvalue $z(k)$. As the damping
coeffcient increases for $k>1$~\AA$^{-1}$, the peak structure of $C^t(k,\omega)$ becomes less pronounced, yielding the corresponding reduction of the well depth in the TCF\index{time correlation functions (TCFs)} $F_{J^{\perp}J^{\perp}}(k,t)$. Though the existence of a negative domain of the transverse ``current-current'' TCF\index{time correlation functions (TCFs)} resembles that in VAFs of the fluids at large densities (see Sec.~5.1), its meaning is obviously different: whereas the VAFs minima describe an essentially single particle cage effect, the minima of TCF\index{time correlation functions (TCFs)}  $F_{J^{\perp}J^{\perp}}(k,t)$ correspond to the collective shear waves. However, the physical origin of such specific behaviour 
in the both cases could be similar and is connected with the phenomenon of local ordering in the fluid. At the same time, the charac\-te\-ris\-tic length of the locally ordered domain can be estimated as $1/k_H$.
\begin{figure}
	\includegraphics[width=5.5cm]{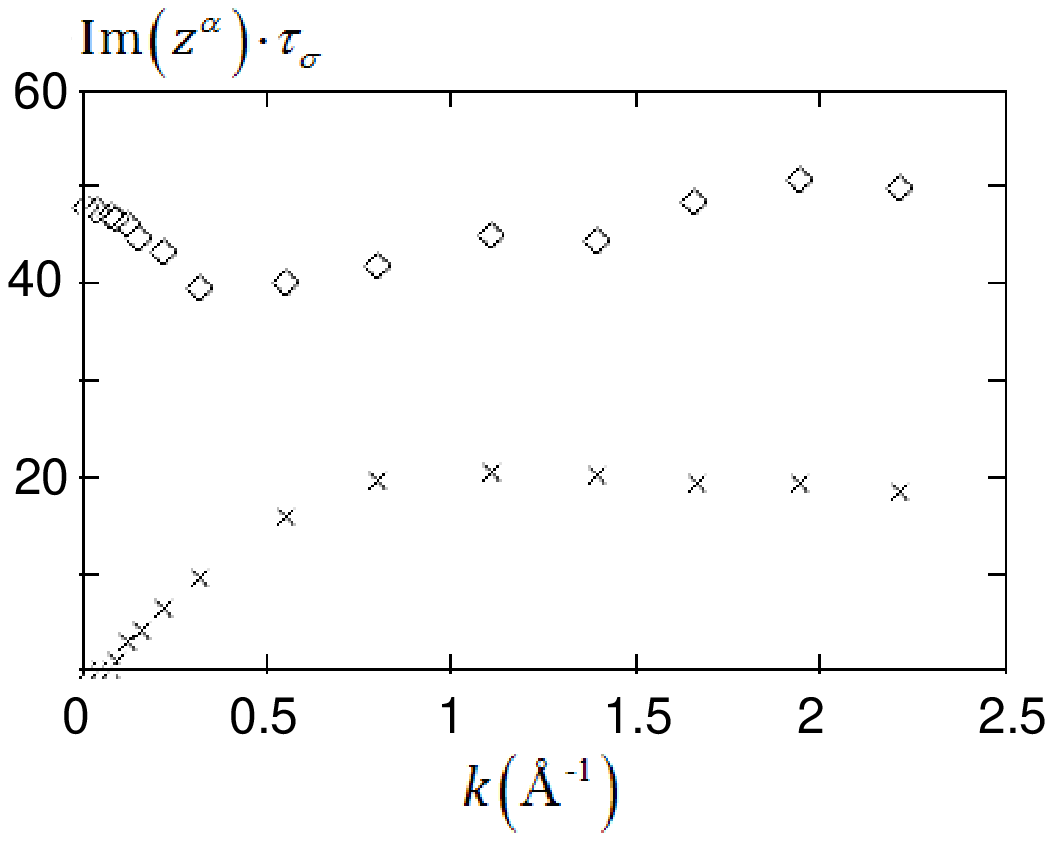}
	\quad
	\includegraphics[width=5.5cm]{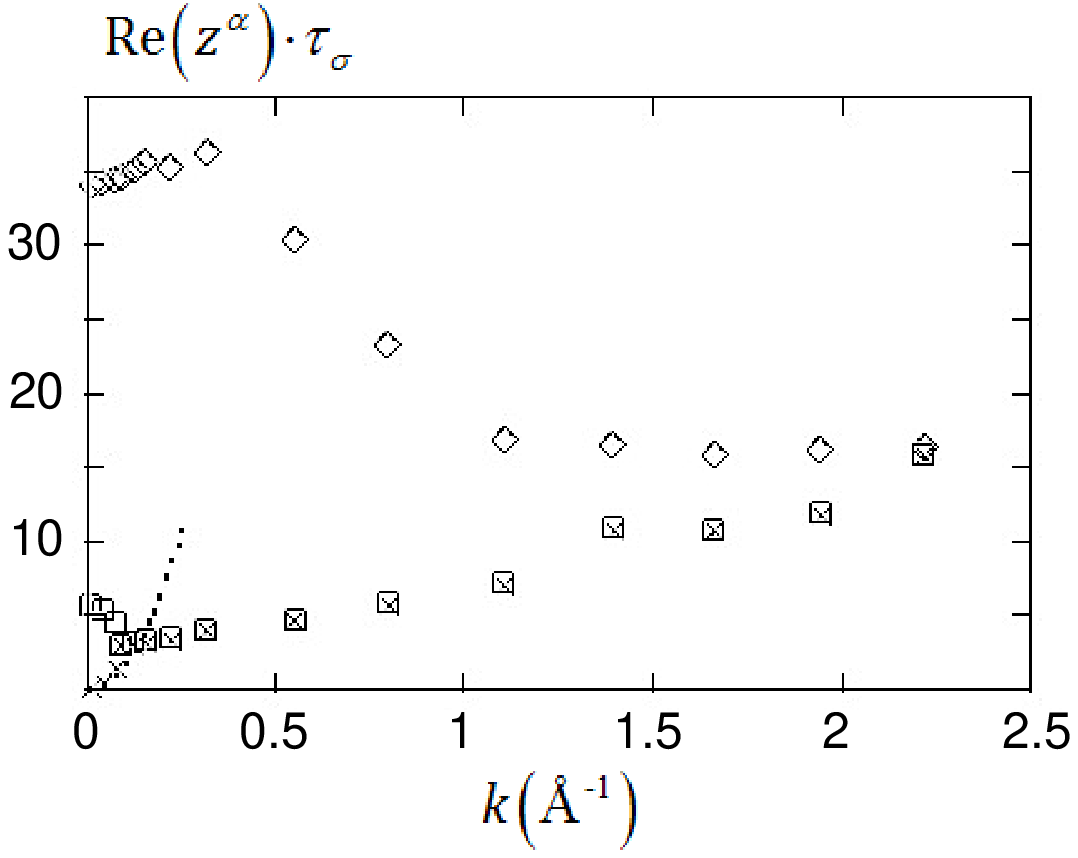}
	\caption{The spectrum of generalized collective modes: 
		imaginary (left panel) and real (right panel) parts of reduced eigenvalues at $M=4$. Dotted curve displays the hydrodynamic behaviour $z^h$. 
	} \label{fig_p38}
\end{figure}

It is also instructive to look at the collective excitation spectrum, presented in Fig.~\ref{fig_p38}. For $k>0.092$~\AA$^{-1}$ the spectrum of transverse excitations consists of two pairs of complex	conjugated propagating modes\index{propagating modes}. The high-frequency and low-frequency modes have comparable real parts (damping) of the eigenvalues only at $k>1$~\AA$^{-1}$. For smaller $k$,  the contribution from the upper mode into all dynamical processes is very small, and in the hydro\-dyna\-mic limit it almost vanishes. At $k<0.092$~\AA$^{-1}$ the lower-lying propagating mode\index{propagating modes} disappears and transforms into two relaxing modes\index{relaxing modes} with purely real eigenvalues. In the $k\to 0$ limit, the eigenvalue of one of these modes tends to a finite damping coefficient, while the second eigenvalue behaves as $\mbox{Re} \,z^h=\displaystyle\eta k^2 / \varrho$ in a full agreement with the hydrodynamic theory. It should be also emphasized that in highly viscous fluids as well as in glass-like systems \cite{BrMr98} with large shear viscosity coefficient the range of hydro\-dyna\-mic behaviour is very small ($k_H \sim 1/ \eta$), so that such fluids behave like elastic bodies up to the characteristic length $\sim 1/k_H$. 

One has to mention that from the mathematical viewpoint the model (\ref{eqn-transverse}) is rather general and can be used to study  many other cases related to the dynamics of conserved quantities. For instance, within a simi\-lar two-variable model the formation of thermal waves in fluids can be explained taking into account the coupling between the heat density and the density of heat current \cite{h-waves,JPCM2000}. That indicates a certain universality of the above described scenario at the heat waves formation, optic-like modes in mixtures (see Sec.~5.3), spin waves \cite{OMF2001} in magnetic liquids, etc.

Before we start considering other kinds of propagating modes, let us conclude Sec.~5.2 by one more demonstration of the viscoelastic behaviour in fluids. 
\begin{figure}[h]
	\centerline{\includegraphics[width=7cm]{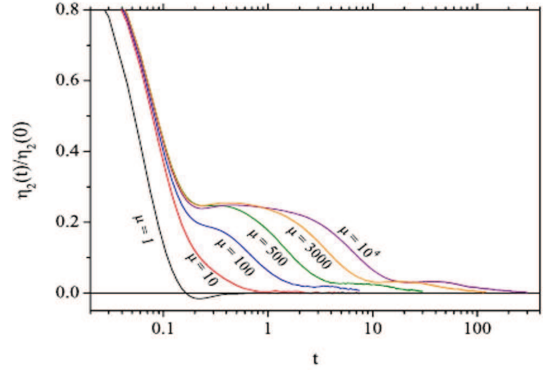}}
	\caption{
		Normalized ``stress-stress'' autocorrelation
		function $\eta_2(t)$ for the heavy component of the binary system with concentration $x=0.2$ of the light species,  dimensionless density $\varrho=0.6$ and temperature $T=1.05$, respectively \cite{FMPF}.
	} \label{fig_p39}
\end{figure}
In Fig.~\ref{fig_p39} we plot the ``stress-stress'' autocorrelation function of the heavier component (labelled here by 2) 
\begin{equation}\label{SAF}
\eta_2(t)=\left\langle\sigma_{xy}^{(2)}(t)\sigma_{xy}^{(2)}(0)\right\rangle
\end{equation}
of the binary Lennard-Jones mixture studied in Ref.~\refcite{FMPF}. 

It is obvious that with increasing mass ratio $\mu$, the relaxation times\index{relaxation time} grow very similarly to the VACF of the heavy component $\psi_2(t)$ (see the right panel of Fig.~\ref{fig_p35}). However, unlike $\psi_2(t)$, the ``stress-stress'' autocorrelation function exhibits  a distinct plateau, which can be considered as a forerunner of the glass phase formation. Such a delayed relaxation of $\eta_2(t)$ anticipates the large values of the fluid viscosity\index{viscosity}, which is expressed by the Green-Kubo relation from the corresponding ``stress-stress'' autocorrelation function. So one can conclude that the strongly asymmetric binary mixture\index{binary mixtures} behaves in a very glassy-like manner, as far as  its viscosity\index{viscosity} is concerned. Obviously,  more complicated dynamic models are needed to describe such behaviour.

The last issue, which we consider in Sec.~5.2, is a possibility to observe the non-Newtonian\index{non-Newtonian fluid} behaviour in simple fluids. Using the two-variable model (\ref{eqn-transverse}) developed within the linear response theory \cite{NSO3,MrHach95}, one can relate the mean value of the transverse stress tensor $\langle\sigma_T\rangle^{\omega}$ to the strain $\gamma_T(\omega)$ as follows:
\begin{equation}\label{sigmaT}
\langle\sigma_T\rangle^{\omega}=i\omega\eta(\omega)\gamma_T(\omega),
\end{equation}
where $\eta(\omega)$ is the generalized shear viscosity\index{viscosity} that has the Lorenz-like form
\begin{equation}\label{eta-Lorentz}
\eta(\omega)=\frac{\eta}{i\omega\tau+1}
\end{equation}
with the corresponding relaxation time\index{relaxation time} $\tau=\eta/G$. The expressions (\ref{sigmaT})-(\ref{eta-Lorentz}) allow one to write down
\begin{eqnarray}\label{NH-laws}
\langle\sigma_T\rangle^t\simeq \eta \frac{\partial \gamma_T(t)}{\partial t}&\quad \mbox{for times}& \quad t\gg\tau,\\
\nonumber
\langle\sigma_T\rangle^t\simeq G \gamma_T(t)&\quad \mbox{for times}& \quad t\ll\tau.
\end{eqnarray}
Therefore, at large times the fluid behaves as a Newtonian liquid\index{Newtonian fluid} (the stress is defined by the strain rate), whereas at small times the Hooke's law is obeyed (see the second equation in (\ref{NH-laws})), and the fluid dynamics resembles that of the elastic body.

\subsection{Optical phonon-like excitations and fast sound}

Let us consider now some problems typical for the collective dynamics of fluid mixtures. The most interesting questions that arise in this case are the following: is there really a strong resemblance between the collective excitations in fluids and the phonon dynamics in crystals?  Could the so-called ``optic-like collective excitations'' be observed experimentally? 

It is well known that the optic-like phonon  excitations in solids describe opposite motions of particles in different species. So, if the cage effects are well pronounced in the fluid, and the nearest neighbours of a particle are formed mainly by the particles of different species, one may hope to find such collective modes in experiments. Otherwise, from the standard hydrodynamics we know that in the hydrodynamic limit only one pair of propagating modes\index{propagating modes} exists in a fluid mixture, and these modes are associated with acoustic sound excitations describing (like in crystals) the coherent movements of all particles. How should the theory be modified in order to treat this problem in more details? 

We start our treatment from the case of binary mixtures\index{binary mixtures}. For this purpose we introduce a new dynamic variable defined as a normal coordinate to the density of total current $\hat{\mathbf J}_{t,\mathbf k}=\hat{\mathbf J}_{\mathbf 1, k}+\hat{\mathbf J}_{2,\mathbf k}$, defined as a sum of its partial components. A new dynamic variable can be introduced via the mass-concentration current according to
\begin{equation}\label{Jx}
\hat{\mathbf J}_{x,\mathbf k}=x_2 \hat{\mathbf J}_{\mathbf 1, k}-x_1 \hat{\mathbf J}_{\mathbf 2, k}.
\end{equation}
Note that $x_a=m_a N_a/\sum_{a} m_a N_a$ with $a=1, 2$, and the new variable $\hat{\mathbf J}_{x,\mathbf k}$ is orthogonal to 
$\hat{\mathbf J}_{t,\mathbf k}$ in the sense that the corresponding SCF\index{static correlation functions (SCFs)} is equal to zero, 
$\left(\hat{\mathbf J}_{t,\mathbf k},\hat{\mathbf J}_{x,\mathbf k} \right)=0$. This means that these dynamic variables are weakly coupled at the equilibrium, and some simplified dynamic models can be used for the subsequent analysis.

Thus using the two-variable model $\hat{\mathbf P}_x^{\perp}=\left\{\hat J_{x,\mathbf k}^{\perp},\dot{\hat J}_{x,\mathbf k}^{\perp}
\right\}$, it is straightforward to find the condition (see Sec.~5.2), when the transverse propagating mass-concentration waves appear. In the hydrodynamic limit this condition looks as follows \cite{Bryk2000}:
\begin{equation}\label{deltaXT}
\delta_x^{\perp}=\frac{\bar{\omega}^{\perp}_{2,J}(0)D_{12}^2 S_{xx}^2(0)\bar m^2}{4 (x_1 x_2 k_B T)^2}<1,
\end{equation}
where $\bar{\omega}^{\perp}_{2,J}(0)=\lim\limits_{k\to 0} \left\langle\dot{\hat J}_{x,\mathbf k}^{\perp}\dot{\hat J}_{x,-\mathbf k}^{\perp}\right\rangle / \left\langle\hat J_{x,\mathbf k}^{\perp}\hat J_{x,-\mathbf k}^{\perp}\right\rangle $ denotes the normalized second-order frequency moment; $D_{12}$ means the mutual diffusion coefficients; $S_{xx}(0)$ denotes the mass-concentration static structure factor at $k=0$, and $\bar m=\sum_a m_a c_a$, $c_a=N_a/N$.

In the long-wavelength limit a similar condition has been derived \cite{Bryk2002} for the longitudinal dynamics of binary mixtures\index{binary mixtures} 
within the three-variable model with the set of dynamic variables $\hat{\mathbf P}_x^{||}=\left\{\hat n_{x,\mathbf k},
\hat J_{x,\mathbf k}^{||},\dot{\hat J}_{x,\mathbf k}^{||} \right\}$. In this case, in addition to the pair of the propagating excitations, the appearance of which is described by a condition like (\ref{deltaXT}), there exists a purely diffusive mode related to the mutual diffusion process. In the generalized version, when the $k$-dependence is taken into account, this condition can be written down in the form 
\begin{equation} \label{deltaXL}
\delta_x^{||}(k)=\frac{\bar{\omega}^{||}_{4,x}(k)}{[2\tau_{xx}(k)\bar{\omega}^{||}_{2,x}(k)]^2}<1,
\end{equation}
where 
\begin{eqnarray*}
	\bar{\omega}^{||}_{2,x}(k)=\frac{k^2 \left\langle {\hat J}_{x,\mathbf k}^{||} {\hat J}_{x,-\mathbf k}^{||}\right\rangle}{\bar{m} \left\langle\hat n_{x,\mathbf k}^{||}\hat n_{x,-\mathbf k}^{||}\right\rangle},\quad \bar{\omega}^{||}_{4,x}(k)=\frac{\left\langle\dot{\hat J}_{x,\mathbf k}^{||}\dot{\hat J}_{x,-\mathbf k}^{||}\right\rangle}{\left\langle\hat J_{x,\mathbf k}^{||}\hat J_{x,-\mathbf k}^{||}\right\rangle},
	\\ \tau_{xx}(k)=S_{xx}^{-1}(k)\int\limits_0^{\infty}F_{xx}(k,t),
\end{eqnarray*}
denote the normalized second and fourth order frequency moments respectively, while $\tau_{xx}(k)$ is the corresponding relaxation time, determined by the mass-concentration TCF\index{time correlation functions (TCFs)} $F_{xx}(k,t)$.

\begin{figure}
	\centerline{\includegraphics[width=6cm]{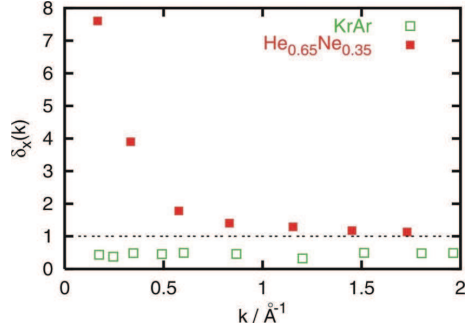}}
	\caption{
		$\delta^{||}_x(k)$ as a function of the
		wave\-number for KrAr and He$_{0.65}$Ne$_{0.35}$. optic-like excitations exist in the long-wavelength
		region only when $\delta_x^{||}(k) < 1$. The results are obtained in Ref.~\refcite{Bryk2002}.
	} \label{fig_p47}
\end{figure}
The condition (\ref{deltaXL}) has been verified for several binary mixtures\index{binary mixtures}, and some results, obtained \cite{Bryk2002}, in particular, for KrAr and He$_{0.65}$Ne$_{0.35}$ are shown in Fig.~\ref{fig_p47}. It is seen that in the hydrodynamic limit this condition is fulfilled for the first mixture, but not for the second one, so that an additional pair of propagating modes\index{propagating modes}, describing the collective opposite motion of particles in different species, could be expected in the KrAr mixture for small $k$.
\begin{figure}[h]
	\centerline{\includegraphics[width=6cm]{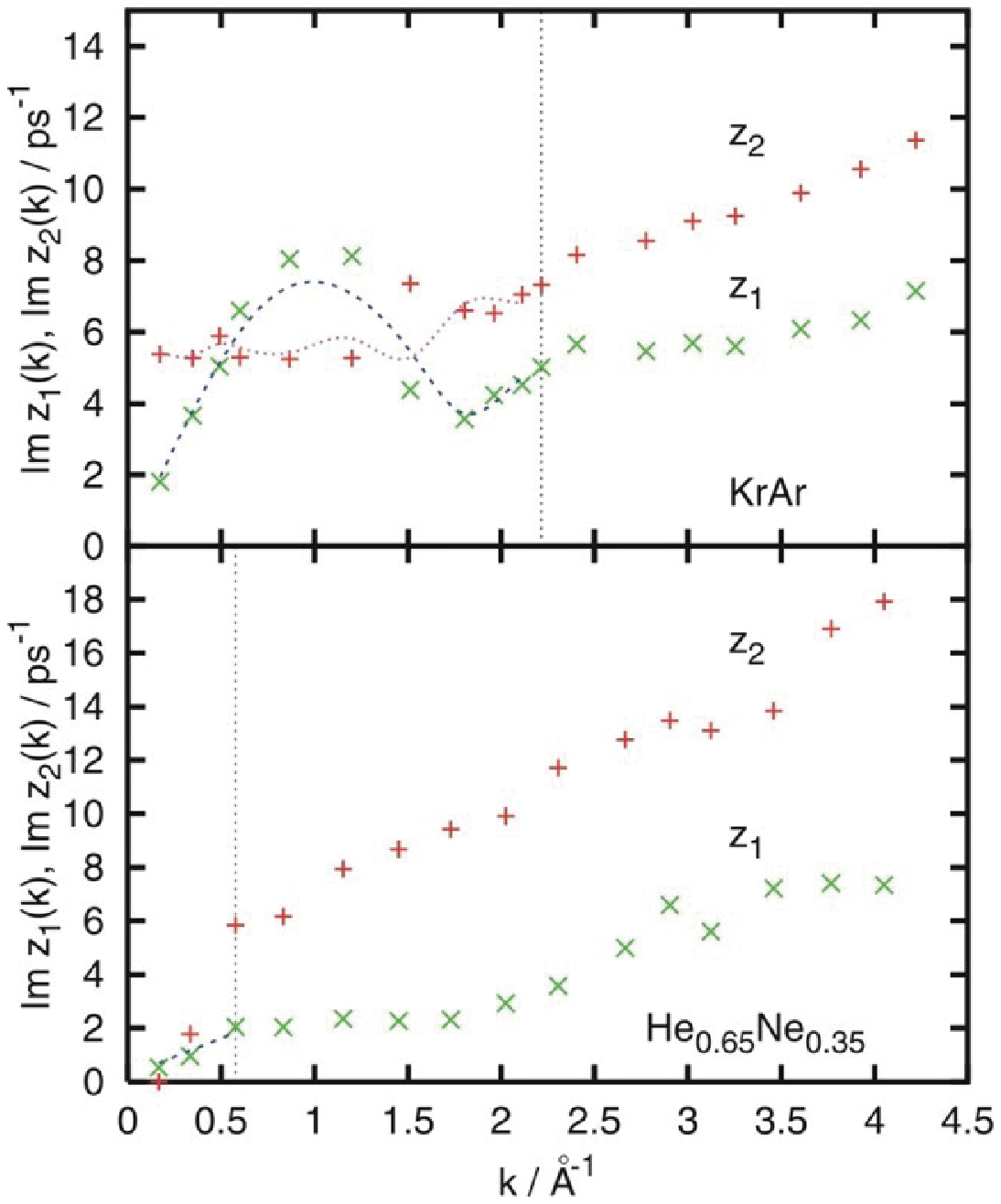}}
	\caption{
		Dispersions of the sound ($z_1(k)$) and high-frequency ($z_2(k)$) excitations in the equimolar
		Lennard-Jones mixture KrAr at 116 K and the dense gas mixture He$_{0.65}$Ne$_{0.35}$ at 39.3 K,
		calculated numerically within the 14-variable GCM approach. Dashed and solid curves
		correspond to dynamical eigenmodes obtained by separate treatment of the dynamics using the 5-variable 
		sets with labels $x$ and $t$, correspondingly. Dotted vertical lines approximately separate the 
		short-wavelength regions of ``partial'' behaviour of branches. The branch $z_2(k)$ corresponds in the 
		long-wavelength region to optic-like excitations in KrAr, while for He$_{0.65}$Ne$_{0.35}$ at 39.3 K 
		the optic-like excitations are suppressed\cite{Bryk2002}.
	} \label{fig_p48-2}
\end{figure}

Reliability of the GCM\index{generalized collective mode (GCM)} approach for the analysis of mass–concentration fluctuations is illustrated in Fig.~\ref{fig_p48-2}, where the imaginary parts of the corresponding propagating eigenvalues for the equimolar mixture KrAr and the dense gas mixture with disparate masses He$_{0.65}$Ne$_{0.35}$ are plotted. 
The spectra of the longitudinal collective excitations
were obtained within the numerical parameter-free 14-variable GCM\index{generalized collective mode (GCM)} approach \cite{Bryk2002}, which included: i) total (labelled $t$) and mass-concentration (labelled $x$) number densities, ii) the corresponding current densities, and iii) total energy density along with the derivatives of variables ii) and iii) up to the third order. All the ele\-ments of the 14$\times$14 generalized hydrodynamic matrix $T(k)$ were evaluated directly in MD\index{molecular dynamics (MD)} simulations, so no free para\-meter was invoked. The number of dynamical variables was chosen to be in agreement with the nine-variable basis set, used for the case of simple li\-quids \cite{GCM1} with the time derivatives of the hydrodynamic variables up to the third order, which allowed one to obtain a very accurate description of the LJ fluid. 

Let us analyse the results for the mode dynamics of both systems. The branch $z_1(k)$ corresponds to the hydrodynamic sound excitations with the linear dispersion in the small wavenumber region. The branch $z_2(k)$ in the long-wavelength
limit is caused by mass–concentration fluctuations and can be reproduced by treatment of solely dynamical variables describing the mass–concentration fluctuations.

\begin{figure}
	\centerline{\includegraphics[width=6cm]{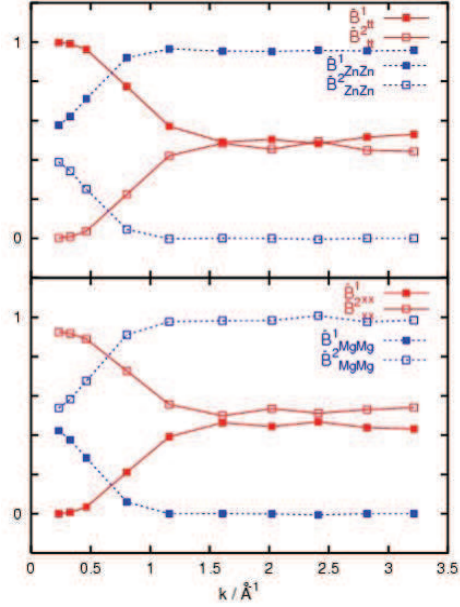}}
	\caption{The normalized weighting coefficients $\bar B_{ij}^p(k)$ of Mg$_{70}$Zn$_{30}$  for the  two lowest
		propagating excitations $z_1(k)$ (closed boxes) and $z_2(k)$ (open boxes) for four different TCFs. Solid and dashed lines in the upper frame correspond to the contributions
		to the autocorrelation functions $F_{tt}(k,t)$ and $F_{\mbox{\tiny{ZnZn}}}(k,t)$, respectively. In the lower frame the solid
		and dashed lines correspond to the  functions $F_{xx} (k,t)$ and $F_{\mbox{\tiny{MgMg}}}(k,t)$, respectively\cite{Bryk2000}.
	} \label{fig_p50}
\end{figure}

The vertical dotted lines in Fig.~\ref{fig_p48-2} separate two regions of wavenumbers, in which either the collective (small $k$) or partial (large $k$) forms of dynamics prevail (see also Fig.~\ref{fig_p50} and the subsequent discussion). One can see in Fig.~\ref{fig_p48-2} that in the case of the dense gas mixture He$_{0.65}$Ne$_{0.35}$  the short-wavelength region of ``partial'' dynamics begins at much smaller wavenumbers than for KrAr. Moreover, in the long-wavelength limit, in contrast to the branch patterns in KrAr, there exists only one branch of propagating sound excitations $z_1(k)$ with the linear dispersion, while the branch $z_2(k)$ is suppressed at $k\to 0$. The dashed and solid curves in Fig.~\ref{fig_p48-2} indicate that the collective dynamics of the KrAr mixture in the short-wavelength domain can be well described by the five-variable sets, built on the variables i) and ii), while the influence of the energy fluctuations is not essential. At the same time, this is not the case for the He$_{0.65}$Ne$_{0.35}$ system, since the region with the ``collective'' dynamics is too narrow to draw any conclusion about the mode behaviour within the five-variable models.

To analyse the collective vs. partial behaviour in the binary mixtures\index{binary mixtures}, it is useful to go back to the transverse dynamics. In the general case within the GCM\index{generalized collective mode (GCM)} approach the equation for the TCF\index{time correlation functions (TCFs)} $F_{ij}(k,t)$ can be written as
\begin{equation}\label{19inBryk2000}
\frac{F_{ij}(k,t)}{F_{jj}(k)}=\sum\limits_{r=1}^{n_r}\bar{A}_{ij}^{r}e^{-\sigma_r t}+\sum\limits_{p=1}^{n_p}\left[
\bar{B}_{ij}^{p}\cos(\omega_p t)+\bar{C}_{ij}^p\sin(\omega_{p} t)\right]e^{-\sigma_p t},
\end{equation}
where the first summation runs over all of the relaxing modes\index{relaxing modes}  $n_r$, while the second one is performed over all the propagating modes\index{propagating modes} $n_p$; $\bar{A}_{ij}^r(k)$, $\bar{B}_{ij}^p(k)$, and $\bar{C}_{ij}^p(k)$ denote the corresponding weighting coefficients; $\omega_p(k)$ means the dispersion of the propagating mode\index{propagating modes}, while $\sigma_r(k)$, $\sigma_p(k)$ are the damping coefficients of the relaxing and propagating excitations, respectively. This result can be obtained within the GCM\index{generalized collective mode (GCM)} approach for the dynamic model with the set of $n_r+2n_p$ dynamic variables.

The interplay between the ``partial'' and ``collective'' behaviour of TCFs\index{time correlation functions (TCFs)} is well seen in Fig.~\ref{fig_p50} on the example of the eight-mode transverse dynamics in a glass-forming molten alloy Mg$_{70}$Zn$_{30}$ at $T$ = 833~K, $n$ = 0.0435~\AA$^{-3}$. The study has been performed \cite{BrMr98,Bryk2000} within the GCM\index{generalized collective mode (GCM)} approach for the eight-variable model based on the partial densities of transverse momentum and its time derivatives up to the 3-rd order. For the  wavenumbers $k > 0.05$~\AA$^{-1}$ we found four branches of propagating modes\index{propagating modes}, i.e. $n_r$ = 0 and $n_p = 4$. The obtained results \cite{Bryk2000} allowed us to calculate the weighting coefficients (see (\ref{19inBryk2000})) for several transverse TCFs\index{time correlation functions (TCFs)} being interest of. The $k$-dependences of normalized amplitudes $\bar B_{ii}^p(k)$ with $i = \{t, x, \mbox{Mg}, \mbox{Zn}\}$ for the lowest two propagating modes\index{propagating modes} (that make the main contributions and are related to shear-waves and optic-like excitations) are plotted in Fig.~\ref{fig_p50}. As one can see, for large $k$  the partial TCFs\index{time correlation functions (TCFs)} $F_{\mbox{\tiny{ZnZn}}}(k, t)$ and $F_{\mbox{\tiny{MgMg}}}(k, t)$ are almost completely determined by the branches $z_1(k)$ and $z_2(k)$, respectively. The same can be said about the contributions of these modes to the TCFs\index{time correlation functions (TCFs)} $F_{tt}(k, t)$ and $F_{xx}(k, t)$ at small $k$. Thus at small $k$ the collective transverse dynamics dominates, whereas the domain of large wavenumbers favours the ``partial'' dynamics. The crossover region is well observed nearby $k \simeq 0.7$~\AA$^{-1}$. 

\begin{figure}
	\includegraphics[width=5.5cm]{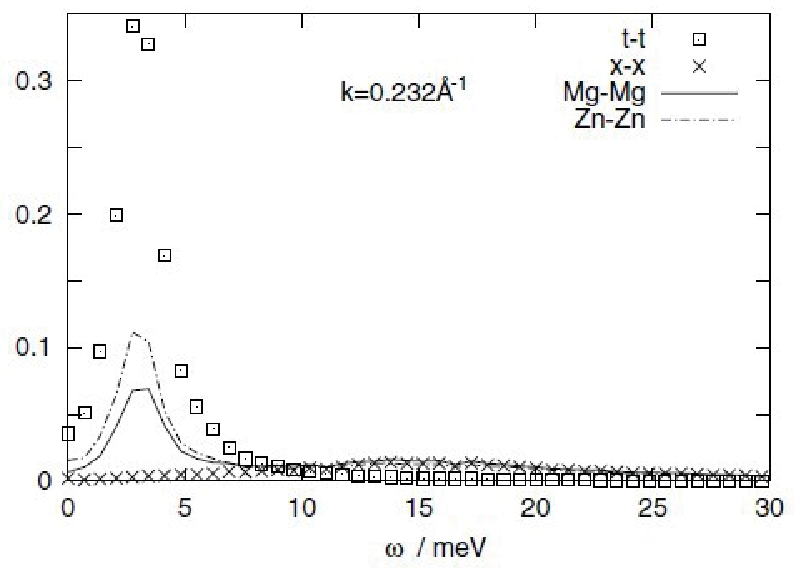}
	\quad
	\includegraphics[width=5.5cm]{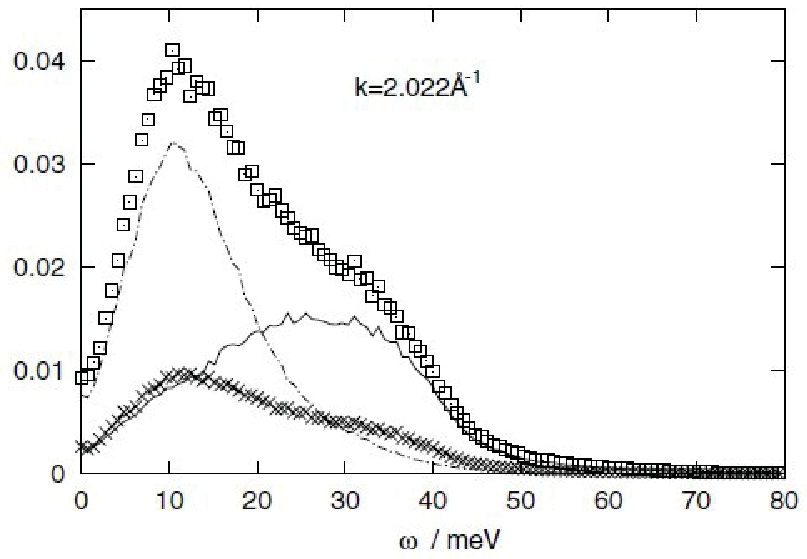}
	\caption{The spectral functions $F_{tt}(k,\omega)$ (open boxes), $F_{xx}(k,\omega)$ (crosses), $F_{\mbox{\tiny MgMg}}(k,\omega)$ (solid line), and $F_{\mbox{\tiny ZnZn}}(k,\omega)$ (dashed–dotted line) for Mg$_{70}$Zn$_{30}$ at two different values of  $k$ \cite{Bryk2000}. 
	} \label{fig_p44}
\end{figure}
In Fig.~\ref{fig_p44} the transverse spectral functions  $F_{tt}(k,\omega)$, $F_{xx}(k,\omega)$, $F_{\mbox{\tiny MgMg}}(k,\omega)$, and $F_{\mbox{\tiny ZnZn}}(k,\omega)$ for Mg$_{70}$Zn$_{30}$, obtained as numerical Fourier transforms of the relevant MD-derived TCFs\index{time correlation functions (TCFs)}, are shown for two values of $k$ that belong to the regions with dominant collective and partial types of collective behaviour, respectively. One can see that the spectral functions of the dynamical variables $J_{t,\mathbf k}^{\perp}$ and $J_{x,\mathbf k}^{\perp}$ for $k=$0.232 \AA$^{-1}$ have the one-peak structure with well-defined maxima, positions of which correspond closely to the frequencies of collective excitations obtained by the GCM\index{generalized collective mode (GCM)} method. The partial spectral  functions for the smaller value of $k$ have the first peak located nearly at the frequency of the shear wave branch, which is much more pronounced in both partial spectral functions than the shoulder  (or a heavily smeared peak) associated with the optic-like excitations with larger damping coefficient. For the larger $k$ = 2.022 \AA$^{-1}$, the situation is quite the opposite: the partial spectral functions manifest a one-peak structure, while the functions $F_{tt}(k,\omega)$ and $F_{xx}(k,\omega)$ each exhibit a main peak (nearly at the position of the maximum for $F_{\mbox{\tiny ZnZn}}(k,\omega)$ and a shoulder (close to the position of the maximum for $F_{\mbox{\tiny MgMg}}(k,\omega)$). This is in a complete agreement with our discussion of the mode contributions and the results presented in Fig.~\ref{fig_p50}.

The ideas, verified for the case of binary fluids, can be developed and applied for the study of propagating optic-like modes in a many component mixture. In the case of a ternary mixture the set of the orthogonalized dynamic variables can be defined in the form \cite{Bryk2007}
\begin{eqnarray}\label{J-ternary}
\nonumber
&&\hat J_{t,\mathbf k}^{\perp}=\hat J_{A,\mathbf k}^{\perp}+\hat J_{B,\mathbf k}^{\perp}+\hat J_{C,\mathbf k}^{\perp},\\
&&\hat J_{x1,\mathbf k}^{\perp}=x_A\hat J_{B,\mathbf k}^{\perp}-x_B\hat J_{A,\mathbf k}^{\perp},\\
\nonumber
&& \hat J_{x2,\mathbf k}^{\perp}=(x_A+x_B)\hat J_{C,\mathbf k}^{\perp}-x_C(\hat J_{A,\mathbf k}^{\perp}+\hat J_{B,\mathbf k}^{\perp}),
\end{eqnarray}
where $x_\alpha$ with $\alpha = A, B, C$ are the corresponding mass-concentrations. In order to analyse a crossover from the collective to partial dyna\-mics, the ternary Lennard-Jones mixture with the molar ratio 2:1:1 and mass ratio 13.91:8.63:4.63 at temperature $T = 116$~K and density $n$=0.0182~\AA$^{-3}$ has been studied within the GCM\index{generalized collective mode (GCM)} approach \cite{Bryk2007}. 
\begin{figure}
	\centerline{\includegraphics[width=8cm]{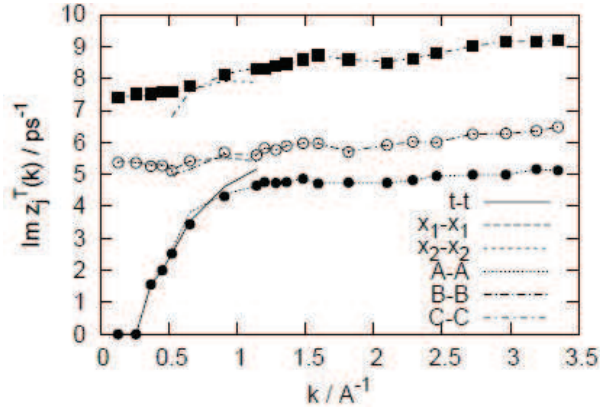}}
	\caption{Imaginary parts of the transverse complex-conjugated eigenvalues, obtained for the
		six-variable basis set $\hat{\mathbf P}^{(6\perp)}$ (symbols) and calculated on the separated subsets of dynamical
		variables (lines) $\hat{\mathbf P}^{(2\perp)}_i$ and $\hat{\mathbf P}^{(2\perp)}_{\alpha}$
		for small and intermediate $k$, respectively\cite{Bryk2007}.
	} \label{fig_p52}
\end{figure}
The collective mode spectra were calculated for the six-variable basic set $\hat{\mathbf P}^{(6\perp)}=\{\hat J_{A,\mathbf k}^{\perp},\hat J_{B,\mathbf k}^{\perp},\hat J_{C,\mathbf k}^{\perp},\dot{\hat J}_{A,\mathbf k}^{\perp}, \dot{\hat J}_{B,\mathbf k}^{\perp},\dot{\hat J}_{C,\mathbf k}^{\perp} \}$ as well as several separated subsets of dynamical variables, namely for  $\hat{\mathbf P}^{(2\perp)}_{\alpha}=\{\hat J_{\alpha,\mathbf k}^{\perp}, \dot{\hat J}_{\alpha,\mathbf k}^{\perp}
\}$ with $\alpha=\{A,B,C\}$ and $\hat{\mathbf P}^{(2\perp)}_{i}= \{\hat J_{i,\mathbf k}^{\perp}, \dot{\hat J}_{i,\mathbf k}^{\perp}\}$ with $i=\{t,x1,x2\}$. The obtained results for the imaginary parts of the transverse complex-conjugated eigenvalues are shown in Fig.~\ref{fig_p52}. The crossover range nearby $k \simeq$0.8~\AA$^{-1}$ is well seen, such that: i)~for larger $k$ the partial dynamics dominates and the separated partial subsets $\hat{\mathbf P}^{(2\perp)}_{\alpha}$ match perfectly the results for the six-variable dynamical model; ii) at smaller wavenumbers the results for imaginary parts of the transverse complex conjugate eigenvalues are well reproduced with the help of the two-variable subsets $\hat{\mathbf P}^{(2\perp)}_{\alpha}$ (see (\ref{J-ternary})), describing the collective behaviour. Thus we conclude that the crossover from the collective behaviour to partial one in fluid mixtures can be observed not only in the $k$-dependences of weighting coefficients (see Fig.~\ref{fig_p50}), but also in 
the $k$-dependences of imaginary parts of propagating collective modes.

Let us return briefly to the case of binary fluids with a significant mass difference of particles. Two important consequences of this difference are usually observed: i) the crossover line to the partial behaviour is shifted to smaller wavenumbers and ii) the propagating modes\index{propagating modes} are strongly separated at larger $k$. It is well seen in Fig.~\ref{fig_p48-2} for the He$_{0.65}$Ne$_{0.35}$ mixture with the mass ratio 5.04. 
\begin{figure}
	\centerline{\includegraphics[width=6cm]{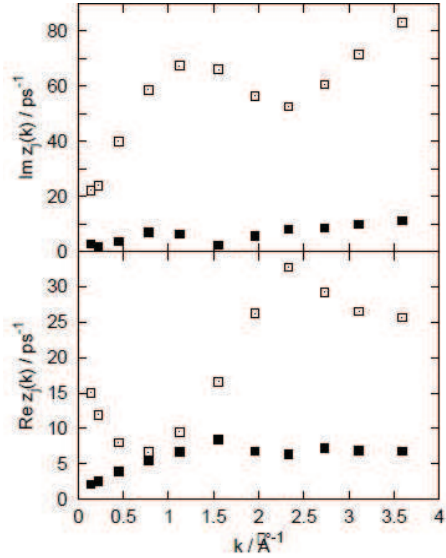}}
	\caption{The imaginary and real parts of the longitudinal complex eigenvalues $z_j(k)$ corresponding
		to the frequency and damping of the propagating collective excitations, respectively \cite{Bryk2004}.
	} \label{fig_p49}
\end{figure}

In Fig.~\ref{fig_p49} the results, obtained in Ref.~\refcite{Bryk2004} for the generalized propagating collective modes of the Li$_4$Pb  molten alloy with mass ratio 29.85 at the temperature of 1085~K and mass-density of 3556.76~kg/m$^3$ are presented. 
As can be seen from the top part of Fig.~\ref{fig_p49}, in the long-wavelength region, the frequency of the concentration propagating mode\index{propagating modes} decreases sharply to the values typical to sound excitations, but its attenuation coefficient increases. For larger wavenumber $k$, the dispersion curves of the generalized concentration and sound modes are strongly separated, and the regime of partial behaviour is observed. The closeness of the frequencies of the propagating modes\index{propagating modes} in the region of small $k$ with the subsequent rapid growth of the difference between them creates good preconditions to explore these excitations in binary mixtures\index{binary mixtures} with large mass ratio in light scattering experiments. One can trace this issue back to the 1980ies, when the first investigations of the phenomenon, named later as the fast sound\index{fast sound}, were performed \cite{Bos86} for the Li$_4$Pb  molten alloy and compared to the prediction of the Mori-Zwanzig formalism. A new propagating mode\index{propagating modes} was found to appear at $k\sim 0.1$~\AA$^{-1}$, showing the linear dispersion in the wavenumber region 0.1~\AA$^{-1} \le k \le$~0.6~\AA$^{-1}$ and having the velocity, which is more than three times higher than that of the ordinary sound. The results of our studies (see also \cite{8inBryk2004,Bryk2004,crossover}) support the conclusion of Ref.~\refcite{Bos86} about the possibility to observe a higher-frequency propagating mode\index{propagating modes} in a binary system of two species with a large mass difference. Moreover, we establish that such propagating excitations are in fact the propagating mass-concentration modes, which behave in a rather specific way in a binary fluid with the large mass ratio. We note that in most cases studied for the binary fluids with large mass ratio, it has been found that the frequency of propagating mass-concentration mode tends to zero at some fixed and very small $k$ (like it was observed for the frequency of the generalized transverse modes discussed in Sec.~5.2). 

For another large group of binary mixtures\index{binary mixtures}, it was found that the frequency of propagating mass-concentration modes in the long-wavelength limit tends to a finite non-zero value, so one has a reason to talk about the optic-like excitations. The question about the possibility of its experimental observation is discussed in more detail for the case of molten salts\index{molten salt}.

\subsection{Dynamics of molten salts: optic-like excitations}

From the theoretical point of view the main difference in the description of molten salts in comparison with binary mixtures\index{binary mixtures} of neutral particles is the nature of the inter-particle interactions. Since a molten salt\index{molten salt} can be considered as a system of charged particles or ions, the interactions between them are described by the long-range Coulomb potential. Besides, in a molten salt\index{molten salt} the electroneutrality condition  $\sum_{\alpha} q_\alpha n_\alpha = 0$ should be satisfied, and this imposes certain restrictions on the concentrations of ions of different charges. In this case, it is more convenient to use a dynamic variable that is associated with the charge density $\hat{\mathbf J}_{q,\mathbf k}$ instead of its mass-concentration counterpart (\ref{Jx}), which is usually used for the binary mixtures\index{binary mixtures} of neutral particles. Because of the electroneutrality condition one has
\begin{equation}\label{Jq}
\hat{\mathbf J}_{q,\mathbf k}=q_1 \hat{\mathbf J}_{\mathbf 1, k}+ q_2 \hat{\mathbf J}_{\mathbf 2, k}= \Bigg[\frac{q_1}{m_1}-\frac{q_2}{m_2}\Bigg]\hat{\mathbf J}_{x,\mathbf k}.
\end{equation}
Thus one can use our previous results obtained for the binary mixture\index{binary mixtures} of neutral particles, in particular, the conditions (\ref{deltaXT}) and (\ref{deltaXL}). Due to the long-range character of Coulomb interaction, one has  the condition $S_{qq} (k)\sim k^2$ for the  ``charge-charge'' SCF\index{static correlation functions (SCFs)} in the hydrodynamic limit. This means that the conditions (\ref{deltaXT}) and (\ref{deltaXL}) are always fulfilled at small wavenumbers $k$. Therefore, the optic-like propagating modes\index{propagating modes} always exist in molten salts. 

\begin{figure}
	\centerline{\includegraphics[width=6cm]{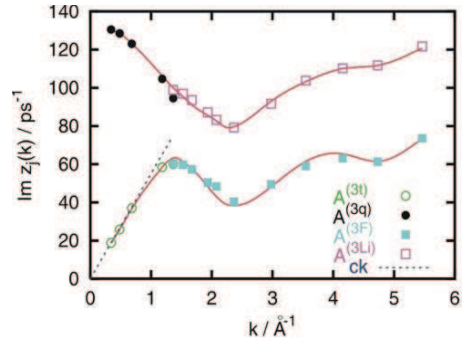}}
	\caption{ The dispersion of propagating collective excitations in molten LiF at 1287 K obtained in
		the eight-variable model (spline-interpolated solid curves). Symbols are the results obtained for
		separate sets of three dynamical variables. The dashed line shows the linear dispersion for sound
		excitations with $c = 5320$~m/s\cite{JPCM2004}.
	} \label{fig_p53-1}
\end{figure}

In order to illustrate this statement and to discuss some specific features of the collective excitations in molten salts\index{molten salt}, the results obtained for the liquid LiF system in Ref.~\refcite{JPCM2004} are displayed in Fig.~\ref{fig_p53-1}.
The collective modes analysis was performed for the eight-variable basic set, including the mass and charge densities, mass and charge current densities, energy densities as well as derivatives of the last three dynamical variables. Besides, the four separate sets $\hat{\mathbf P}^{(3i)}_{\mathbf k}=\left\{\hat n_{i,\mathbf k}, \hat J^{||}_{i,\mathbf k}, \dot{\hat J}^{||}_{i,\mathbf k} \right\}$, $i=\{m, q, \mbox{Li},\mbox{F} \}$ (the labels $t$ and $q$ correspond to mass and charge, respectively), were also taken into consideration to analyse the collective modes spectra in more details.

In Fig.~\ref {fig_p53-1} the imaginary parts of two complex eigenvalues $z_{\alpha}(k)$, which correspond to
the propagating excitations, are shown. Another branch of propagating excitations, corresponding to heat waves, was obtained in the region $k > 0.8$~\AA$^{-1}$; however, here we focus on the propagating density fluctuations only. The dispersion curves shown by solid spline-interpolated lines were estimated from the eight-variable model. One can conclude that in the region $k < 1.4$~\AA$^{-1}$ the high-frequency branch is solely defined by pro\-pa\-ga\-ting charge waves, while the low-frequency branch comes from the fluctuations of the total mass density and in the long-wavelength region shows an almost linear dispersion law with the propagation speed $c = 5320$~m/s. In this region the low- and high-frequency branches correspond to acoustic and optical phonon-like excitations, respectively. 
For larger wavenumbers $k > 1.4$~\AA$^{-1}$, the partial behaviour of both branches was established, with the low and high-frequency branches describing solely the heavy (F) and light (Li) subsystems in the melt. Like in the case of non-ionic mixtures (cf. Fig.~\ref{fig_p52}), the separate three-variable sets, labelled by $t$ and $q$, perfectly fit the data of the eight-variable model at small $k$, whereas at large wavenumbers the results obtained using the reduced basic sets, labelled by F and Li, show a good agreement with those of the extended eight-variable model. Another difference from the non-ionic mixtures is that the collective excitations in the molten LiF salt are well separated in frequency at all $k$, while for the mixture of neutral particles it is not the case (see Fig.~\ref{fig_p48-2} for comparison).
\begin{figure}
	\centerline{\includegraphics[width=7cm]{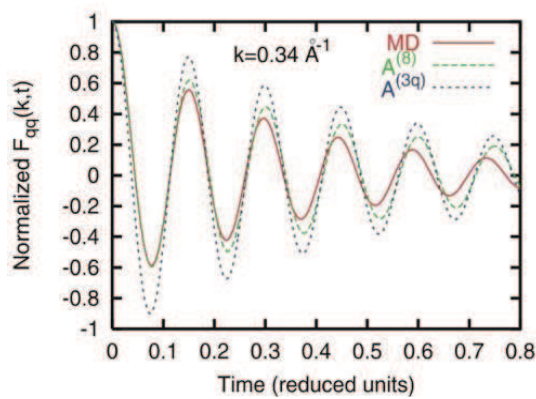}}
	\caption{ The charge density autocorrelation function for molten LiF obtained in MD\index{molecular dynamics (MD)} simulations
		(solid curve) and its GCM\index{generalized collective mode (GCM)} replicas estimated from the eight-variable (dashed curve) and
		three-variable (dotted curve) models\cite{JPCM2004}.
	} \label{fig_p53-2}
\end{figure}

In Fig.~\ref{fig_p53-2} we show two GCM replicas for the smallest wavenumber charge density autocorrelation function, obtained with
the extended eight-variable set of dynamic variables (dashed curve) and the reduced three-variable one (dotted curve), no fitting parameters in both cases. The GCM replica obtained from the three-variable treatment of the charge subsystem reproduces the oscillating behaviour of $F_{qq} (k, t)$, but contains oscillations that are less overdamped, because the interaction
between charge fluctuations and other hydrodynamic processes was not taken into account. In the case of the eight-variable treatment, the GCM data fit the MD results at small and intermediate times well enough.

As it has been already mentioned, one of the advantages of the GCM\index{generalized collective mode (GCM)} approach is the possibility to separate the mode contributions to the GCM replica, see, for instance, Eq.~(\ref{19inBryk2000}).
\begin{figure}
	\centerline{\includegraphics[width=7cm]{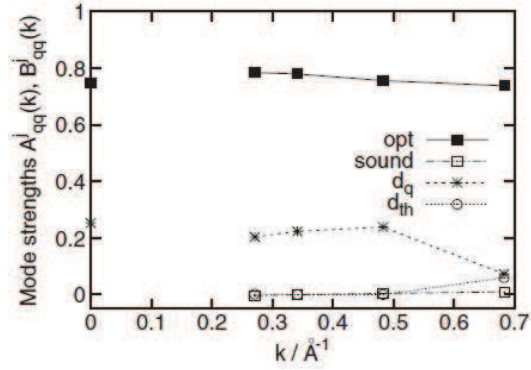}}
	\caption{ Mode amplitudes of different relaxing and propagating modes\index{propagating modes} for the charge density
		autocorrelation functions obtained for molten LiF in the eight-variable GCM model (symbols
		connected by lines)\cite{JPCM2004}. 
	} \label{fig_p53-3}
\end{figure}
In Fig.~\ref{fig_p53-3} the mode amplitudes for the two main relaxation processes of the mutual diffusion $A^q_{qq}(k)$, the thermal diffusivity $A^{th}_{qq}(k)$ as well as the two symmetric  contributions coming from optical and acoustic branches of collective excitations are shown by symbol-connected lines for the case of eight-variable model of collective dynamics in molten LiF. It is seen that for $k < 0.5$~\AA$^{-1}$ neither the sound excitations nor the thermal diffusivity contribute to the ``charge–charge'' TCF\index{time correlation functions (TCFs)} $F_{qq}(k, t)$, and the shape of $F_{qq}(k, t)$ is determined solely by the relaxation process of the electric conductivity and propagating charge waves. Moreover, the contribution from the
non-hydrodynamic charge waves is almost four times as large as that from the hydrodynamic relaxation process. This shows the striking difference between the molten salts\index{molten salt} with the long-range interaction and non-ionic liquid mixtures, for which the mode strength of optic-like excitations in $F_{xx}(k, t)$ vanishes rapidly towards small wavenumbers \cite{Bryk2002}.

With this in mind, let us try to answer the question: is it possible to observe the optic-like excitations in scattering experiments considering the shape of dynamic structure factor? In Ref.~\refcite{crossover} the dynamic properties and the crossover between hydrodynamic and kinetic modes in liquid alloys Na$_{57}$K$_{43}$ were investigated by applying the GCM\index{generalized collective mode (GCM)} ana\-ly\-sis scheme and using the inelastic X-ray scattering (IXS) data. The GCM\index{generalized collective mode (GCM)} scheme was built using the extended eight-variable formalism with $\hat{\mathbf P}^{(8)}=\{\hat n_{t,\mathbf k},\hat n_{x,\mathbf k},\hat J_{t,\mathbf k}^{||}, \hat J_{x,\mathbf k}^{||}, \hat{\varepsilon}_{\mathbf k}, \dot{\hat J}_{t,\mathbf k}^{||}, \dot{\hat J}_{x,\mathbf k}^{||},\dot{\hat\varepsilon}_{\mathbf k}
\}$ as well as the reduced three-variable subsets $\hat{\mathbf P}^{(3\alpha)}=\{\hat n_{\alpha,\mathbf k},\hat J_{\alpha,\mathbf k}^{||}, \dot{\hat J}_{\alpha,\mathbf k}^{||}\}$, $\alpha=\{t, x, \mbox{Na}, \mbox{K}\}$. Usage of the different sets of dynamic variables allows us to ascertain the origin of each branch of collective excitations in the spectrum and processes responsible for their appearance in various wavenumber domains. 

\begin{figure}
	\centerline{\includegraphics[width=8cm]{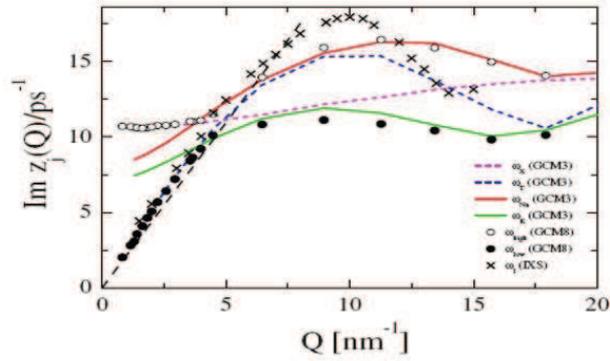}}
	\caption{Experimental dispersion relation \cite{crossover} obtained from the maxima of $C_{\mbox{\tiny {IXS}}}(Q,\omega)$ ($\times$), along with the hydrodynamic dispersion derived by ultrasound measurements \cite{32incrossover} (black dashed line). The two propagating modes predicted by complete eight-variable GCM theory (high frequency, $\circ$; low frequency, $\bullet$), and the outcome of partial
		3-variable GCM analysis with total (blue dashed line), concentration (pink dashed line), Na (red solid line), and K (green solid line) variables are also reported.
	} \label{fig_p51-1}
\end{figure}
The result for spectral current density is reported in Fig.~\ref{fig_p51-1} and clearly shows that the partial type of dynamics (full red and green lines for the Na and K subsets, respectively) is dominant for $Q >5 $~nm$^{-1}$. The corresponding excitations are very close to the two eigenvalues of the eight-variable treatment in this range, while at smaller $Q$ the same eigenvalues are reproduced by the total density and concentration subsets. This is a clear indication of the existence of two dynamical regimes: the collective region at low $Q$ and the domain of partial dynamics at higher $Q$, in a complete agreement with our previous conclusions (see also \cite{5incrossover}).
\begin{figure}
	\centerline{\includegraphics[width=7cm]{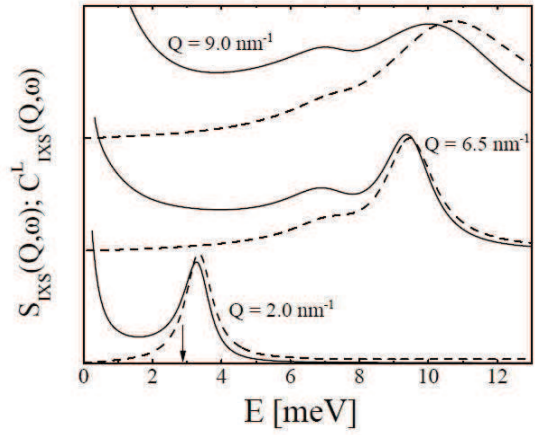}}
	\caption{Classical resolution deconvoluted $S(Q,\omega)$ (continuous line) and $C_L(Q,\omega)$ (dashed lines) obtained in Ref.~\refcite{crossover} from the IXS measurement. The presence of a single mode exceeding the adiabatic frequency (marked by the arrow) is clearly visible at $Q=2$~nm$^{-1}$, i.e. well below the crossover to the partial type dynamics. For $Q=6.5$~nm$^{-1}$ two excitations appears, at the high frequency one is dominating at $Q=9$~nm$^{-1}$.  
	}\label{fig_p51-2}
\end{figure}

The presence of the two phonon-like modes in the IXS spectra at intermediate $k$ can be conveniently quantified
by looking at their relative weights reported in Ref.~\refcite{crossover} and presented in Fig.~\ref{fig_p51-2}. The low frequency mode is clearly dominant below the
sharp crossover occurring at $Q\simeq 6$~nm$^{-1}$. Around this value, both modes contribute to the IXS spectra, while above it the
high frequency mode domi\-na\-tes up to a new crossover at $Q\simeq 13$~nm$^{-1}$. Since the IXS cross section is roughly proportional
to the total density autocorrelation function, below the crossover it reflects the collective longitudinal
excitation and not the optic-like mode. At small $Q$, the relative weight of the optic-like propagating modes\index{propagating modes} at small wavenumbers is proportional to $Q^2$ as for any other collective excitation of the kinetic origin. 

\begin{figure}
	\centerline{\includegraphics[width=7cm]{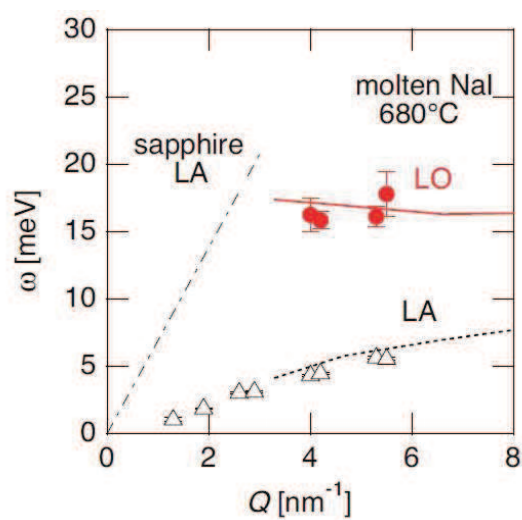}}
	\caption{Dispersions of the collective propagating excitations in molten NaI. Circles and triangles indicate the experimental data for the longitudinal optic (LO) and acoustic (LA) modes, respectively, and the solid and dashed curves show the theoretical ones \cite{11inBryk2010}.	 
	}\label{fig_add_2}
\end{figure}

Similar results have been obtained in the recent paper~\cite{CMP2019}, where the high-resolution IXS measurements were carried out on the molten NaI near the melting point at 680$^0$C. The measured spectra agreed well (in both frequency and linewidth) with the \textit{ab initio} molecular dynamics simulations but not with the classical ones. The observation of these modes at small $k$ and a good agreement (see Fig.~\ref{fig_add_2}) with the simulations permits their clear identification as the collective optic-like excitations with the well defined phasing between different ionic motions. Besides, the obtained data for the dispersion relations and spectra linewidth fit perfectly into the theoretical curves obtained within the GCM approach \cite{11inBryk2010}.

Though the dynamic properties of ionic liquids and neutral mixtures are similar and can be studied within the same theoretical and experimental approaches, there are some peculiarities dealt with a non-zero contribution of the optic-like excitations to the ``charge-charge'' dynamic structure factor even in the hydrodynamic limit. 
Fortunately, in the case of the mixtures with the long-range Coulomb interaction between their particles, the problem of observation of the optic-like excitations can be solved even experimentally. A theoretical description of these excitations in ionic liquids was a subject of numerous studies, and up to now some controversies in the obtained results  are not explained.

\subsection{Some rigorous relations and their applications}

Let us return now to the expressions for the elements of memory functions (\ref{PhiNN})-(\ref{PhiCross}) that look rather sophisticated for practical applications. One of the possibilities to proceed is to use them for calculation of the generalized transport coefficients (see, for instance, Ref.~\refcite{Mry97c}). But there is also another option that allows us to derive some rigorous relations being useful in practice. 

\begin{figure}
	\centerline{\includegraphics[width=7cm]{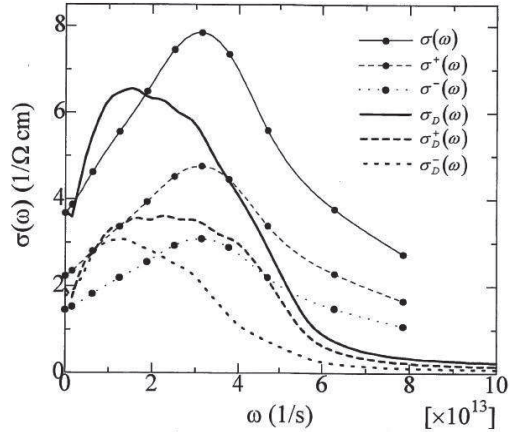}}
	\caption{The frequency dependences of electrical conductivities in molten NaI calculated in Ref.~\refcite{Koishi} by using MD (lines with symbols) and within the theory using $\omega$-dependent self-diffusion coefficients (lines). 
	}\label{fig_add_3}
\end{figure}

One of the most interesting results, important in the context of our consideration of molten salts\index{molten salt}, was obtained in the middle of the XIX century by Johann Wilhelm Hittorf. He reasonably has pointed out that some ions travel more rapidly than others, and this observation led to the concept of the transport number, the fraction of the electric current carried by each ionic species. He measured the changes in the concentration of electrolyzed solutions, computed from these transport numbers (relative carrying capacities) of the ions, and established his laws governing the migration of ions. A century later, B.~R.~Sundheim demonstrated \cite{Sundheim} that simple considerations of the conservation of electrolyte particles momentum lead to definite predictions about the relative motions of parts of such ionic complexes as measured in their particular rearrangement. It has been shown that the transference numbers $t_{\alpha}$ of the species in a pure molten salt\index{molten salt} can be expressed in terms of the masses $M_{\alpha}$ as $t_{\alpha}=M_{\alpha}/\sum_{\nu} M_{\nu}$. Since every considered electrolysis cell has to obey the condition of electroneutrality $t_+q_+-t_-q_-=0$, and the electric current $J_{\alpha}$ is dealt with the electric field strength $E$ by the Ohm's law, $J_{\alpha}=\sigma_{\alpha} E$, one can anticipate the simple relation between the partial electric conductivities $\sigma_{\alpha}$,
\begin{equation}\label{rel-sigma}
\frac{\sigma_+}{\sigma_-}=\frac{M_-}{M_+}.
\end{equation}
It was obvious that a simple phenomenological treatment of the ionic transport in electrolytes did not allow to obtain anything more accurate than (\ref{rel-sigma}). To move further the statistical mechanical theory had to be applied. In Ref.~\refcite{Koishi} such a study with aim to derive rigorously the relation (\ref{rel-sigma}) was performed for the case of molten NaCl and NaI using the Langevin equations for motion of  cations and anions. The partial ionic conductivities were derived in terms of the ``current-current'' TCF as an extension of the Kubo theory. It was shown that the relation (\ref{rel-sigma}) is fulfilled in this approach, and for the total conductivity $\sigma$ the following expression was found: 
\begin{equation}\label{sigma-Koishi}
\sigma=\sigma_++\sigma_-=\frac{e^2}{k_B T}\left(n_+z_+^2 D_+ + n_-z_-^2 D_-\right)(1-\Delta),
\end{equation}
where $e$ means the electron charge, $n_{\alpha}$ and $z_{\alpha}$ denote the $\alpha$-ion ($\alpha=+,-$) number density and valence, respectively, and $D_{\alpha}$ denotes the corresponding self-diffusion constant. The quantity $\Delta$ is treated as the deviation from the Nernst-Einstein relation, being defined by some cation-anion cross-correlation terms, involving the zeroth time moments of the velocity intercorrelation functions. This $\Delta$ correction was evaluated in the MD\index{molecular dynamics (MD)} simu\-lations, and its absolute value turned out to be smaller then unity in both cases of molten NaCl at $T=1100$ K and NaI at $T=950$. However, for NaCl its value was strongly underestimated as compared with the experimental result for $\Delta$, and for NaI it was twice overestimated, whereas the theoretical results for the diffusion constants were consistent with the experimental ones. In Ref.~\refcite{Koishi} the MD\index{molecular dynamics (MD)} frequency dependent partial conductivities were also calculated and their maxima were found to occur almost at the same frequencies both for NaCl and NaI (see Fig.~\ref{fig_add_3})

As follows from the comparison of the corresponding curves in Fig.~\ref{fig_add_3}, the proposed theory does not allow to capture the main features of the frequency dependence of either complete or partial conductivities. From the physical point of view, this is understandable, because the frequency dependences of the self-diffusion coefficients are largely determined by the Einstein frequencies \cite{BMT2003}, which correspond to the average frequency of oscillations of the tagged particle in the environment of others. It is clear that  the frequency of such oscillations will be different for each particle type, because their masses are different, and so are their immediate environments, which indirectly affect the oscillatory motion of the selected particle. Since the maxima of the partial conductivities are localized at the same frequency, it is obvious that this frequency characterizes the oscillatory motion inherent to the system as a whole and cannot be described in one-particle circuits. So what kind of collective motion is this? We will return to this question a bit later.

In Ref.~\refcite{Matsunaga2004} the structural properties of the $\beta$ phase and transport properties of the superionic
$\alpha$, $\beta$, and molten phases\footnote{These phases differ from each other because of the different arrangement of anions, which can be observed in the corresponding pair distribution functions.} of the Ag$_3$SI system were investigated by the MD\index{molecular dynamics (MD)}
simulations, using the Vashishta–Rahman-type potentials. The most important conclusion of this study, being interesting for our consideration, is the new relation proposed for the partial conductivities 
\begin{eqnarray}\label{Matsunaga}
m_{\tiny\mbox{S} }\sigma_{\tiny\mbox{S}}/|z_{\tiny\mbox{S}}|+m_{\tiny\mbox{I}}\sigma_{\tiny\mbox{I}}/|z_{\tiny\mbox{I}}|=m_{\tiny\mbox{Ag}}\sigma_{\tiny\mbox{Ag}}/|z_{\tiny\mbox{Ag}}|.
\end{eqnarray}
This relation can be considered as a generalized version of (\ref{rel-sigma}) for a ternary mixture. Further studies of the pseudo-binary molten salts\index{molten salt}, performed  by Matsunaga and co-authors in Ref.~\refcite{Matsunaga2007} (AgI-AgBr system) and in Ref.~\refcite{Matsunaga2008} (NaCl-KCl system), have shown that the relation (\ref{Matsunaga}) is valid for these ternary alloys. It then became even more important to  prove such ``a golden rule'' expression within the non-equilibrium statistical mechanics approach. 

In Sec.~3 the equations of the generalized hydrodynamics were derived using the NSO\index{non-equilibrium statistical operator (NSO)} method. It allowed us to obtain the expressions for generalized transport coefficients (\ref{PhiNN})-(\ref{PhiCross}). One of the important achievements of the theory is a possibility to prove a number of exact relations for the generalized transport coefficients immediately from the conservation laws. It follows already from the structure of the expressions for the memory functions, defined for the set of hydrodynamic variables and constructed on the corresponding generalized currents. For the generalized currents $I_{{\bf k},\alpha}^d$ related to the partial densities of particles $\{ \hat{n}_{{\bf k},\alpha} \}$, we have the identity 
\begin{equation}  \label{OF-n-pr}
\sum \limits_{\alpha=1}^{\nu} m_{\alpha} I_{{\bf k},\alpha}^d = \sum
\limits_{\alpha=1}^{\nu} \left( \hat J_{{\bf k},\alpha}^{||} -
\frac{m_{\alpha} c_{\alpha}}{\bar{m}} \hat J_{\bf k}^{||} \right)
\equiv 0,
\end{equation}
from which we directly obtain a number of exact relations for the genera\-li\-zed transport coefficients of a $\nu$-component mixture:
\begin{equation} \label{dpr}
\sum \limits_{\alpha=1}^{\nu} m_{\alpha} D_{\alpha \gamma}(k,z)= \sum \limits_{\gamma=1}^{\nu}  m_{\alpha} D_{\alpha}^T(k,z) = \sum \limits_{\alpha=1}^{\nu} m_{\alpha} \zeta_{\alpha} (k,z) \equiv 0.
\end{equation}
Among them, the mutual diffusion coefficients $D_{\alpha \gamma}$ are the most important for our subsequent analysis.  In the case of the $\nu$-component fluid, the matrix of the mutual diffusion coefficients contains $\nu \times \nu$ elements. Taking into account the symmetry $D_{\alpha \gamma}=D_{\gamma \alpha}$, their number is reduced to $\nu \times (\nu+1)/2$ independent elements. And if we also take into account the exact relations given above, we come to the conclusion that only $\nu \times (\nu -1)/2$ independent ele\-ments remain. Thus, in a binary mixture\index{binary mixtures} there is only one independent coefficient, in a ternary fluid its number rises to three, etc.

Let us now consider an ionic system with the charges $q_\alpha$, masses $m_\alpha$, and concentrations $c_\alpha$. In such a system the electroneutrality condition $\sum_\alpha q_\alpha c_\alpha =0$ holds. For the properly normalized mutual diffusion coefficient $D_{\alpha \gamma} = c_\alpha c_\beta \bar{D}_{\alpha \gamma}$, the exact relation for $\bar D_{\alpha \gamma}$ can be rewritten as 
\begin{equation} \label{dpr-1}
\sum \limits_{\alpha=1}^{\nu} m_{\alpha}c_\alpha \bar{D}_{\alpha \gamma}(k,\omega)= \sum \limits_{\gamma=1}^{\nu} \bar{D}_{\alpha
	\gamma}(k,\omega)c_\gamma m_{\gamma} \equiv 0.
\end{equation}
If an external field ${\bf E}_\omega$ is applied to the system, an expression for the ionic conductivity can be derived within the linear response theory. For the partial conductivities one gets:
\begin{equation}\label{sigma-ab}
\sigma_{\alpha}(\omega)=k_0 q_\alpha c_\alpha \sum\limits_\beta
q_\beta c_\beta \bar{D}_{\alpha\beta}(\omega),
\end{equation}
where $k_0$ is a certain coefficient depending only on the temperature and density of the system. If we multiply $\sigma_{\alpha}(x,\omega)$ by $m_\alpha/q_\alpha$ and perform the summation over $\alpha$, the general relation for partial conductivities can be obtained:
\begin{eqnarray} \label{sigma-sum1}
&&\sum\limits_\alpha \frac{m_\alpha}{q_\alpha} \
\sigma_{\alpha}(\omega)= k_0 \sum\limits_\alpha m_\alpha c_\alpha
\left[\sum\limits_\beta q_\beta c_\beta \bar{D}_{\alpha\beta}(\omega)
\right] \nonumber 
\\
&&= k_0 \sum\limits_\beta q_\beta c_\beta \left[ \sum\limits_\alpha
m_\alpha c_\alpha \bar{D}_{\alpha\beta}(\omega) \right]\equiv 0,
\end{eqnarray}
and the identical equality to zero in (\ref{sigma-sum1}) follows from Eq.~(\ref{dpr-1}).

A more complicated case corresponds to the model of the ($\nu+\bar{\nu}$)-component fluid that is composed of $N_\alpha$
ions with charges $q_\alpha$ ($\alpha=1, 2, \ldots, \nu$) and $N_{\bar{\alpha}}$ neutral particles belonging to the
$\bar{\alpha}$-th species ($\bar{\alpha}=1, 2, \ldots, \bar{\nu}$). In this case, Eq.~(\ref{sigma-sum1}) should be rewritten in
the form
\begin{eqnarray} \label{sigma-sum2}
\sum\limits_{\alpha=1}^{\nu} \frac{m_\alpha}{q_\alpha} \
\sigma_{\alpha}(\omega)= k_0 \sum\limits_{\beta=1}^\nu q_\beta
c_\beta \left[ \sum\limits_{\alpha=1}^\nu
m_\alpha c_\alpha \bar{D}_{\alpha\beta}(\omega) \right]
\nonumber \\
= - k_0 \sum\limits_{\beta=1}^\nu q_\beta c_\beta \left[
\sum\limits_{\bar{\alpha}=1}^{\bar{\nu}} m_{\bar{\alpha}}
c_{\bar{\alpha}} \bar{D}_{\bar{\alpha}\beta}(\omega) \right],
\end{eqnarray}
where the relations (\ref{dpr-1}) have been used.

The expression (\ref{sigma-sum2}) can be further simplified if we introduce two new densities, namely the mass density of solvent
$\hat{M}_{\bf k}$ (formed by neutral particles only) and the charge density $\hat{Q}_{\bf k}$:
\begin{equation} \label{sigma-sum2_2}
\hat{M}_{\bf k} = \sum\limits_{\bar{\alpha}=1}^{\bar{\nu}}
m_{\bar{\alpha}} c_{\bar{\alpha}} \hat n_{\bar{\alpha},{\bf k}}\,, \qquad %
\hat{Q}_{\bf k} = \sum\limits_{\alpha=1}^{\nu} q_\alpha
c_\alpha \hat n_{\alpha,\bf k}\,.
\end{equation}
Now we can rewrite (\ref{sigma-sum2}) in the form
\begin{eqnarray} \label{sigma-sum3}
\sum\limits_{\alpha=1}^{\nu} \frac{m_\alpha}{q_\alpha} \
\sigma_{\alpha}(\omega)= - D_{\rm MQ} (\omega),
\end{eqnarray}
where the generalized transport coefficient
\begin{equation}\label{DMQ}
D_{\rm MQ} (\omega)= k_0 \sum\limits_{\bar{\alpha}=1}^{\bar{\nu}}
\sum\limits_{\beta=1}^\nu \ m_{\bar{\alpha}} c_{\bar{\alpha}} \
\bar{D}_{\bar{\alpha}\beta}(\omega) \ q_\beta c_\beta
\end{equation}
describes the diffusive ion-solvent cross-correlations. The relation (\ref{sigma-sum3}) is, in fact, the most general form of the ``golden rule'' identity valid for partial ionic conductivities of classical systems of charged particles in a neutral solvent. As it has been shown this relation follows directly from the conservation law for the total momentum (see (\ref{dpr-1})) and the electroneutrality condition. Obviously, the expressions (\ref{rel-sigma}) and (\ref{Matsunaga}) can be easily obtained from (\ref{sigma-sum3}). 

Returning to the question about the frequency dependences of partial and total conductivities (see Fig.~\ref{fig_add_3}), we note that $\sigma_1(\omega)/\sigma_2(\omega) = |q_1| m_2 / (|q_2| m_1)$ and this allows: i) to explain why the peak for all the conductivities obtained in the MD\index{molecular dynamics (MD)} simulations is localized at the same frequency; ii) to determine the origin of this excitation. The last statement follows from the fact that frequency-dependent conductivities in a molten salt\index{molten salt} can be expressed via the spectral functions of the charge fluxes in the hydrodynamic limit. As it has been shown earlier, at small $k$ the main contribution to these functions gives a collective mode of optical type. 
The cases when the system in a new phase state changes its conducting properties may be of a particular interest. 

\subsection{Dynamic crossover in supercritical fluids}

One of the most attractive fields for the soft matter theory is the dynamics of supercritical fluids. It is well known that upon increasing the substance pressure and temperature beyond its critical point, any thermodynamic discontinuity between the liquid and gas phase disappears and the system is said to be in a fluid state. A character of the local ordering as well as the physical properties of a system are changing significantly herewith. There are some phenomena related to the collective dynamics of fluids in the supercritical region that completely differ from those known for the systems at more moderate conditions. For instance, the dispersion curve of the acoustic excitation in dense liquids is found to deviate from the linear relation $c_s k$ at small wavenumbers either upwards (yielding the so-called \textit{``positive dispersion''} \cite{7inBryk2010}) or downwards (\textit{``negative dispersion''} \cite{8inBryk2010}). 

To date two main mechanisms are recognized to be responsible for the positive dispersion in fluids: nonlocal coupling between hydrodynamic modes, described within the mode-coupling theory (MCT)\index{mode-coupling theory (MCT)} \cite{4inBryk2010}, and local coupling
between acoustic excitations and non-hydrodynamic structural relaxation obtained within the memory function formalism \cite{9inBryk2010}. However, there was no systematic theoretical and experimental study of the density or pressure dependence of positive or negative deviation from hydrodynamic dispersion law, except for the inelastic neutron scattering experiments and subsequent calculations, based on MCT\index{mode-coupling theory (MCT)}  for liquid Ar at 120 K. The MCT\index{mode-coupling theory (MCT)} yields the following expression for the
dispersion law of sound excitations
\begin{equation}\label{5over2}
\omega(k)=c_s k+\alpha_s k^{5/2}+ O(k^{11/4})+\ldots,
\end{equation}
where the coefficient $\alpha_s$ is expressed in a sophisticated way, requiring the knowledge of the explicit  dependences of adiabatic sound velocity $c_s$ and thermal expansion coefficient $\alpha_T$ on the density. The results of Ref.~\refcite{10inBryk2010} have proven the positivity of the prefactor $\alpha_s$; however, little is known about its density dependence. The recent IXS experiments on supercritical Ar \cite{11inBryk2010} resulted in a completely different tendency for the pressure dependence of the positive sound dispersion in the supercritical region. It was observed that the positive dispersion reduces with decreasing density and finally practically vanishes. 	

In this subsection, we present the results of  Ref.~\refcite{Bryk2010}, where the thorough analysis of the sound dispersion along several isothermal lines on the phase diagram of the supercritical Lennard-Jones fluids was performed within the GCM\index{generalized collective mode (GCM)} approach. The analytical expressions, based on the local-coupling mechanism, were obtained and used for the subsequent investigation of microscopic processes, being responsible for the positive (or negative) sound dispersion at the boundary of the hydrodynamic regime.
\begin{figure}[h]
	\centerline{\includegraphics[height=7cm]{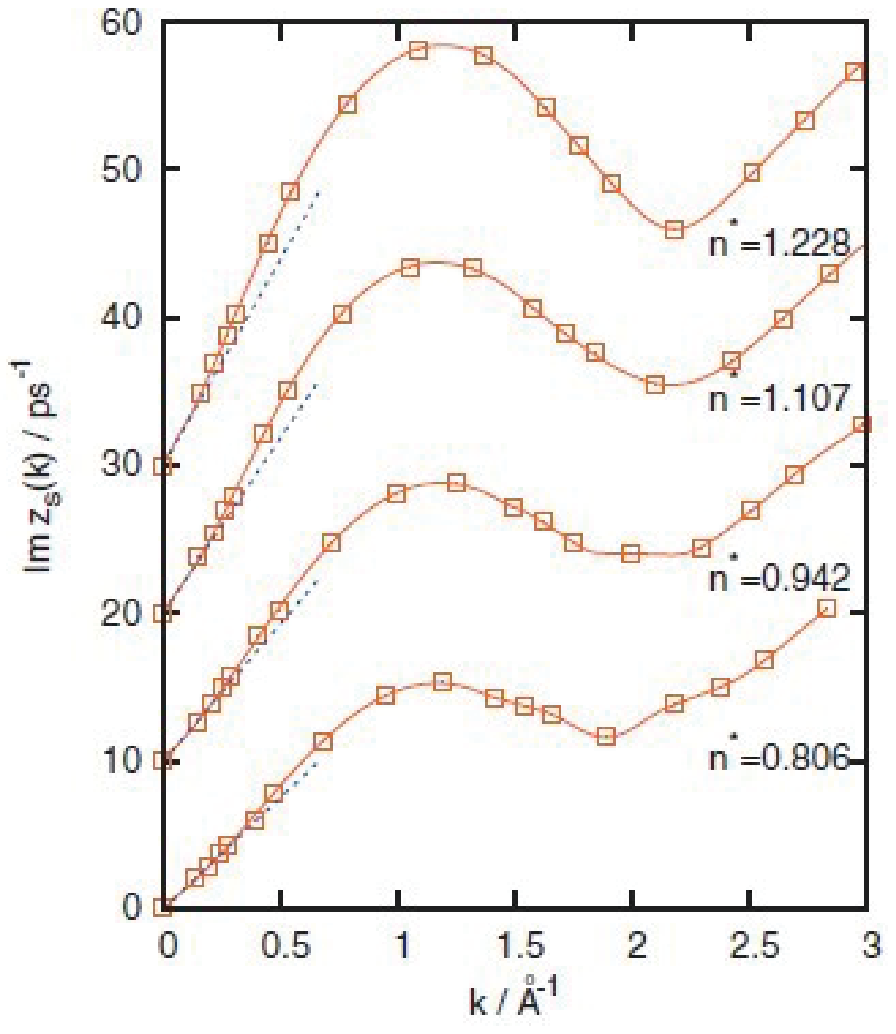}\quad
		\includegraphics[height=7cm]{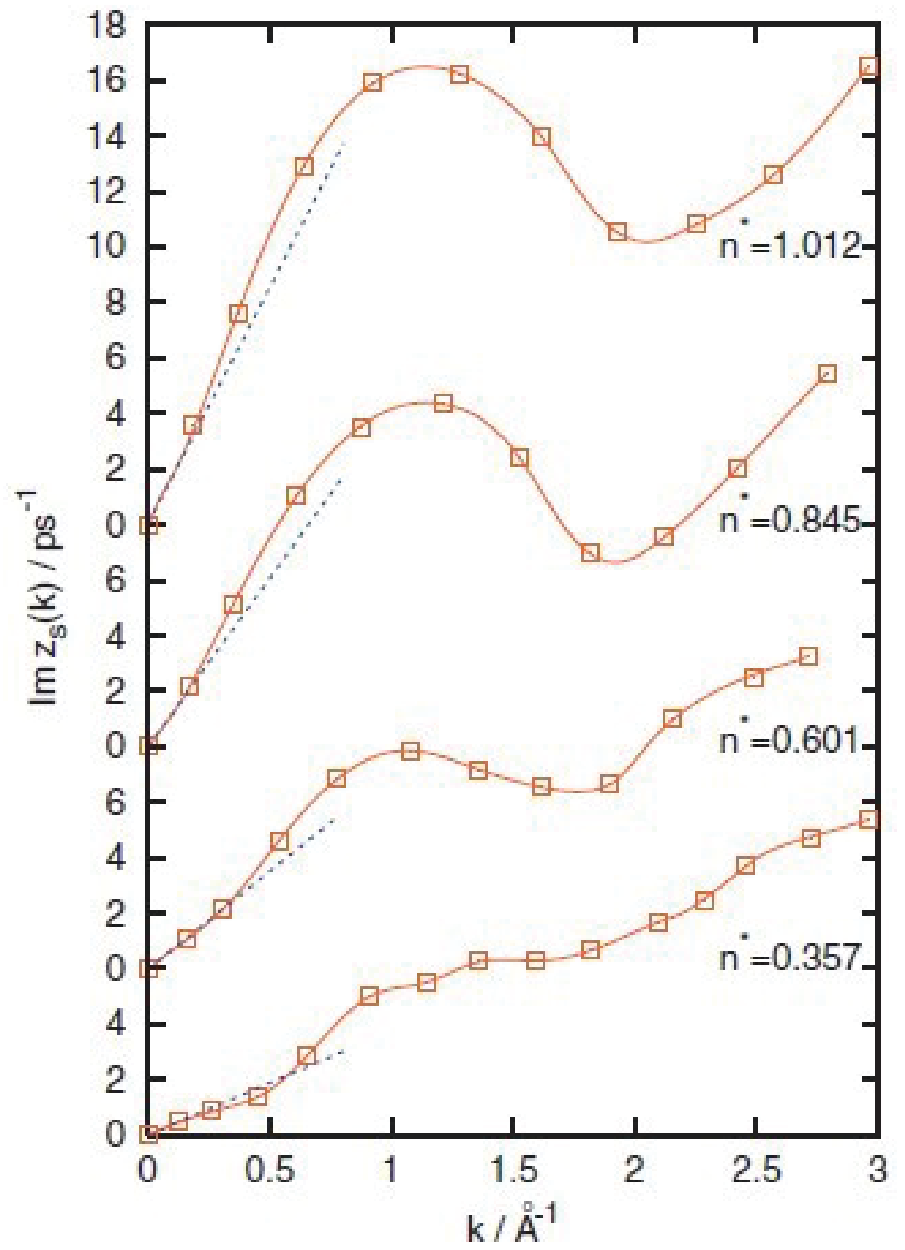}}
	\caption{ Dispersion of collective sound excitations for different densities at $T^*=4.78$ (left) and $T^*=1.71$ (right). Dashed line shows the linear hydrodynamic dispersion law with the hydrodynamic sound velocity $c_s$\cite{Bryk2010}.
	} \label{fig_p42}
\end{figure}

In Fig.~\ref{fig_p42}, the dispersion curves, calculated in Ref.~\refcite{Bryk2010} for several densities, are shown at the reduced temperatures $T^*=4.78$ and 1.71. The general tendencies in the dispersion change with density at $T^*=4.78$ are: i) the reduction of the slope of the linear dispersion law at small $k$ with decreasing density; ii) the reduction of the roton-like minimum with decreasing density; iii) the reduction of positive dispersion with decreasing density; and iv) increasing of the width of region, where the apparent sound velocity almost coincides with $c_s$, with decreasing density.

It is interesting that at some low-density states the eigenvalues with a negative dispersion of collective sound 
excitations are identified in our study. These cases can be found at the right panel of Fig.~\ref{fig_p42} in the
long-wavelength region.

Using the thermo-viscoelastic dynamic model within the GCM\index{generalized collective mode (GCM)} approach the analytical result for the sound dispersion  in the small $k$ domain has been obtained: 
\begin{equation}\label{kcube}
\omega(k)=c_s k+\beta_s k^3+\ldots,
\end{equation}
where the coefficient \\[1mm]
\begin{equation}\label{beta3}
\beta_s =\frac{c_s D_L^2}{8}\,\frac{5-(c_{\infty}/c_s)^2}{c_{\infty}^2-c_s^2}-(\gamma-1)D_T\left[\frac{6D_L+(\gamma-5)D_T}{8 c_s}-\frac{c_s}{2 d_T}
\right]
\end{equation}
is expressed via the adiabatic $c_s$ and high-frequency $c_{\infty}$ sound velocities, the ratio $\gamma$ of specific heats at constant pressure and constant volume, the kinematic viscosity $D_L$, the thermal diffusivity $D_T$, and the long-wavelength limit $k\to 0$ of the kinetic structural relaxation mode $d_T$. Obviously, the expression (\ref{kcube}) gives a different result from  (\ref{5over2}), found by the MCT\index{mode-coupling theory (MCT)}. Note that the first term in (\ref{kcube}) describes the viscoelastic contribution, whereas the second term reflects mainly the thermal processes and vanishes if $\gamma$ tends to zero.

\begin{figure}
	\centerline{
		\includegraphics[width=5.5cm]{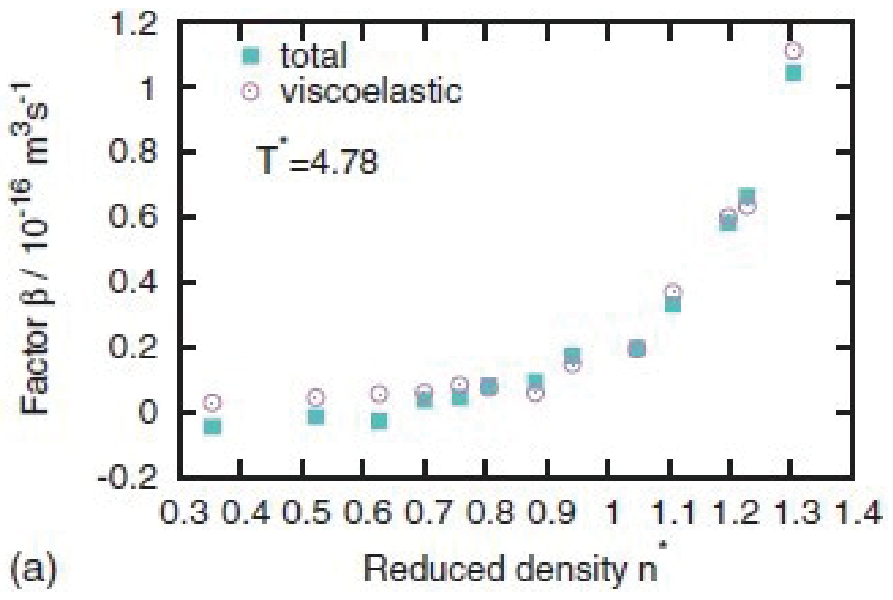}\quad
		\includegraphics[width=5.5cm]{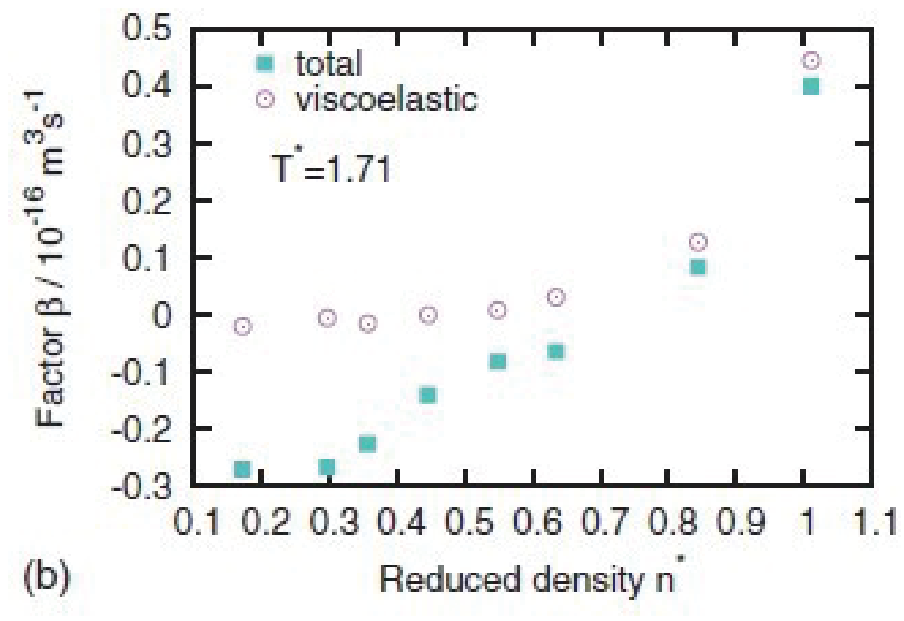}}
	\caption{ Dependence of factor $\beta_s$, Eq.~(\ref{beta3}) on the density for two temperatures
		$T^*=4.78$ and $T^*=1.71$ (closed boxes). Open circles correspond to viscoelastic
		contribution to the $\beta_s$\cite{Bryk2010}.
	} \label{fig_p43}
\end{figure}
In Fig.~\ref{fig_p43}, the results of numerical calculations of the factor $\beta_s$ from Eq.~(\ref{beta3}) at two temperatures
using the GCM data as an input are presented. The strength of the contribution from the first term in (\ref{beta3}) is shown by open circles. The evident difference between the total and viscoelastic contributions is attributed to the thermal processes (the second
term in Eq.~(\ref{beta3})). One can see that for the high-temperature state $T^*=4.78$ the positive dispersion is
completely defined by the viscoelastic mechanism. Most interesting is the fact that the positive dispersion practically vanishes for that temperature around $n^*\approx 0.8$. For lower (and closer to the critical point) temperature $T^*=1.71$, the effect of the thermal contribution is much stronger. At low temperatures for densities $n^*>0.75$ the dispersion curve in the long-wavelength
region shows a positive dispersion, while for $n^*<0.6$ the dispersion can be negative in a narrow region of wavenumbers
close to the hydrodynamic regime. By comparing the results obtained for the factor $\beta_s$ (see Fig.~\ref{fig_p43})
that defines the deviation from linear dispersion law in the long-wavelength region with the dispersion  curves, calculated  within the GCM\index{generalized collective mode (GCM)} approach and shown in Fig.~\ref{fig_p42}, one can conclude that the analytic expression (\ref{kcube}) describes 
the sound dispersion at both densities. Therefore, there is an additional argument to state that the main mechanism responsible for deviation of the sound dispersion from the linear dependence at small $k$ is dealt with the local interaction of collective modes, which is easy to take into account within the GCM\index{generalized collective mode (GCM)} approach. Moreover, the beginning of the range with the subsequent growth of the viscoelastic term in $\beta_s$ (see Fig.~\ref{fig_p43}) can be used as a simple criterion for estimation of the crossover from the gas-like fluid to the liquid-like fluid. A role of the non-hydrodynamic processes in the viscoelastic transition in pure fluids  with the main emphasis on the relation  between the appearance of shear waves in the transverse dynamics and the manifestation of positive sound dispersion in the longitudinal dynamics was studied recently in Ref.~\refcite{PhylMag2020}, but this might be a topic for another review. 

\section{Conclusions}

In this Chapter we have considered some old and new problems in the fluid dynamics, which are interesting from both theoretical and experimental points of view. All of them are united by the same subject of research and the theoretical approach applied. The fluids described in this chapter differ from each other by their constituents (single-component, binary, and ternary systems), types of interactions between the particles (Lennard-Jones, Coulomb and many others), and thermodynamic states considered. At the same time  we tried  to demonstrate that at least some of these problems can be efficiently attacked within the GCM\index{generalized collective mode (GCM)} approach, because: i) it is based on the rigo\-rous theoretical framework that creates the NSO\index{non-equilibrium statistical operator (NSO)} method; ii) it allows to treat effectively the time/space hierarchy of the processes by choosing an appropriate set of dynamical variables; iii) it is non-perturbative and does not use any fitting parameters (the only input data are the SCFs\index{static correlation functions (SCFs)} and the hydrodynamic relaxation times, which can be taken from computer simulations). As the output, one obtains the spectrum of single-particle or collective excitations, which enter the corresponding TCFs additively, and each mode contribution can be traced accurately. At the same time, the results for TCFs\index{time correlation functions (TCFs)} obtained within the GCM\index{generalized collective mode (GCM)}  approach are controlled by the sum rules up to some fixed order and found to be close to those obtained experimentally, in particular in the scattering experiments. 

The extended set of dynamical variables complements the hydrodynamic collective mode spectrum due to the appearance of kinetic modes, which have finite damping coefficients in the hydrodynamic limit. The criteria for these modes to be either purely relaxational or of a propagative nature are formulated. In the latter case, the new phenomena like shear waves or fast sound\index{fast sound} have been predicted and observed both in the experiment and in computer simulations. Sometimes the contributions from the obtained propagative kinetic modes are overlapped by the terms from the generalized hydrodynamic modes. In other circumstances they are well separated from the hydrodynamics mode spectrum and at modest values of the damping can be quite distinguishable at the corresponding TCFs\index{time correlation functions (TCFs)}. We have also pointed out the conditions when the fluid dynamics has a ``collective'' nature and can be best modelled by the mass density and mass concentration variables, or it is of ``partial'' nature, being dominantly described by the densities of particular species of the fluid. The crossover from the collective behaviour to the partial one has a rather general character.

Several simple (but non-trivial) dynamical models have been  considered and analysed in detail. It has been shown that these models can be solved analytically, and the obtained results are very useful for establishing the nature of the generalized collective modes and deriving criteria for their appearance and/or experimental observations. This allows us to study how fluids behave at various time scales and describe the crossover from the dynamics typical for a Newtonian fluid\index{Newtonian fluid} to the elastic solid-like behaviour. It has been shown that even at the level of the single-particle dynamics the collective cage effects can play an important role. For binary fluids the criterium for the appearance of the optic-like propagating modes\index{propagating modes} has been formulated and tested in computer simulations. 

We have also presented some rigorous relations for the transport coefficients of a multicomponent fluid. These relations in the case of a system of charged particles allowed us to derive rigorously the general version of the so-called ``golden rule'' for partial conductivities. It is shown that such a relation can be used as a new opportunity to observe the optic-like modes in a fluid mixture.

Having read the above list of possibilities and merits of the GSM approach, one can ask some natural questions: is this method  suitable for studying the collective dynamics of fluid in the entire space and time domains? What is the relation of the GCM\index{generalized collective mode (GCM)}  method to other theoretical approaches? Is there any limitation of its applicability for the description of fluid dynamics? 

Surely, the collective dynamics of fluids can be best described by those approaches that rest upon strict statistical mechanics relations, which allow generalization of both the thermodynamics and hydrodynamics, and are amenable to computer simulations. In this context, the GCM\index{generalized collective mode (GCM)} approach should be considered as one of the most developed options 
that can be used in combination with others depending on the problem considered.  Still, there remain plenty of questions about its applicability to the dynamics of complex fluids such as solutions of macromolecules and proteins. Hence, answering old question and solving some old puzzles lead to new ones, to be solved soon.



\printindex                        

\end{document}